\documentclass[twocolumn,aps,prl,showpacs,preprintnumbers,amsmath,amssymb,floatfix]{revtex4-2}
\usepackage[utf8]{inputenc}
\usepackage{graphicx}
\usepackage{amsfonts}
\usepackage{bm}
\usepackage{color}
\usepackage{xcolor}
\usepackage{appendix}
\usepackage{mathrsfs}
\usepackage[colorlinks=true,linkcolor=blue,urlcolor=blue,citecolor=blue,anchorcolor=blue]{hyperref}

\newcommand{\dd}{\mathrm{d}}

\newcommand{\KZ}{\mathrm{KZ}}
\newcommand{\AL}{\mathrm{AL}}

\newcommand{\sech}{\operatorname{sech}}
\newcommand{\Ob}{\mathcal O_{\!B}}
\newcommand{\HBCFT}{H_{\mathrm{BCFT}}}

\newcounter{emsection}
\newcommand{\emsection}[1]{%
  \par\medskip
  \refstepcounter{emsection}%
  \setcounter{equation}{0}%
  \renewcommand{\theequation}{\Alph{emsection}\arabic{equation}}%
  \renewcommand{\theHequation}{EM.\Alph{emsection}.\arabic{equation}}%
  \noindent\textit{#1.}---%
}

\begin{document}

\title{Work fluctuation speed limit in boundary conformal field theories}
\author{Shihao Xia, Ahsan Nazir, Harry J. D. Miller}
\affiliation{Department of Physics and Astronomy, The University of Manchester,
Manchester M13 9PL, United Kingdom}
\date{\today}

\begin{abstract}
We explore the fundamental limits on finite-time driving in quantum critical systems  described by boundary conformal field theory. We show that stochastic work fluctuations arising from external driving are a resource for speedy control, and derive an exact, saturable fluctuation-based speed limit in weakly
driven boundary conformal field theories at finite temperature. The bound and saturating protocol can be expressed entirely in terms of the universal scaling dimension, and the result interpolates between the  Kibble--Zurek regime,
where temporal correlations are strongly nonlocal, and an adiabatic regime
where linear driving becomes optimal. For small scaling dimension, the enhanced
temporal correlations produce pronounced departures from linear protocols and a larger
optimization advantage. These
results establish a universal work precision--time tradeoff for boundary-critical
control, applicable to quantum impurity, fractional quantum Hall, and
superconducting-circuit platforms.
\end{abstract}

\maketitle

\textit{Introduction.---} At a boundary quantum critical point 
correlations and relaxation scales diverge, 
strongly constraining finite-time driving and giving rise to universal
nonequilibrium scaling, such as the Kibble--Zurek (KZ) mechanism
\cite{Kibble1976,Zurek1985,Dziarmaga2005,Polkovnikov2005}.
From the perspective of quantum thermodynamics, such driving also generates a
distribution of work whose fluctuations quantify the precision and
irreversibility of the protocol
\cite{Talkner2007,campisi2011colloquium,batalhao2014experimental,cerisola2017using,funo2019quantum}. This raises a natural optimisation question: given a prescribed amount of work
fluctuations, how rapidly can the system be driven away from a boundary critical point and which protocol attains
this minimum time? Here we will answer this question by deriving a work fluctuation-based speed limit for critical systems described by boundary conformal field theories.

The interplay between out-of-equilibrium control, fluctuations and precision has been explored through many different approaches such as quantum speed limits \cite{del2013quantum,deffner2017quantum,funo2017universal}, thermodynamic speed limits \cite{funo2019speed,van2021geometrical,Funo2025,Hasegawa2023,Yamauchi2025} and uncertainty relations \cite{hasegawa2021thermodynamic,miller2021thermodynamic,van2023thermodynamic}. This has been accompanied by the development of a wide range of optimal control methods for non-equilibrium dynamics with applications to quantum thermodynamics \cite{cavina2018optimal,abiuso2020geometric,brandner2020thermodynamic,erdman2023pareto} and shortcuts to adiabaticity \cite{acconcia2015shortcuts,campbell2017trade,guery2019shortcuts}. Finite-time control near quantum criticality has been studied from several
of these perspectives, including many-body state preparation \cite{doria2011optimal}, counterdiabatic driving
\cite{del2012assisted} and quantum speed limits around critical points \cite{heyl2017quenching}. In quantum thermodynamics, 
work statistics and finite-time performance can inherit universal critical
scaling in such regimes \cite{Fei2020,b2020universal,naze2022kibble,Ma2025}.

The overarching class of critical systems that we will consider are those described by boundary conformal
field theory (BCFT). Here conformal invariance fixes the finite-temperature
correlations of boundary operators in terms of their scaling dimensions
\cite{Cardy1989,AffleckLudwig1991,Affleck1995,Cardy2004}. BCFT provides the
low-energy description of a broad range of quantum impurity systems, including
multichannel Kondo fixed points, Luttinger-liquid impurities, and experimentally
realized charge-Kondo devices
\cite{NozieresBlandin1980,CoxZawadowski1998,KaneFisher1992,
Iftikhar2015,Iftikhar2018}. Once the driven boundary operator is specified,
its scaling dimension $\Delta$ therefore determines the universal
finite-temperature response kernel up to overall nonuniversal scales.

In this paper we derive for the first time a saturable bound on the minimum time needed to weakly drive a boundary critical system out of equilibrium, given a prescribed tolerance in work fluctuations at a finite temperature. While problems such as minimising dissipation and finding shortcuts to adiabaticity in weakly-driven systems have been considered previously \cite{bonancca2014optimal,acconcia2015shortcuts,naze2022optimal,Naze2023}, the optimisation of quantum work fluctuations in linear response regimes, let alone boundary critical quantum systems, has yet to be addressed. By modeling such systems as a boundary conformal field theory, our speed limit is characterised entirely through a universal scaling dimension. Near criticality, long-lived correlations can make finite-time work
fluctuations sensitive to the history of the drive, leading naturally to a
nonlocal memory kernel.  Outside the adiabatic regime, optimizing
the fluctuations therefore requires a nonlocal variational problem over the
protocol velocity. We solve this by using a conformal mapping to transform this optimal control problem into a known Riesz energy problem for singular power-law kernels. In turn we have thus found a rare example of an exactly solvable optimal control problem in non-equilibrium quantum dynamics.

\textit{Setup.}---We consider a boundary quantum critical system initially prepared in thermal equilibrium at the boundary fixed point. Its time-dependent Hamiltonian is
\begin{equation}
H_t=\HBCFT+\lambda_t\Ob(x=0),\qquad
\lambda_t=\alpha g(t).
\label{eq:driven-bcft-H}
\end{equation}
Here $x$ denotes the spatial coordinate normal to the boundary; we take the system to occupy $x\geq0$, so $x=0$ is the physical boundary or impurity location in the unfolded description at which the drive acts. The operator $\Ob(x=0)$ is the  boundary scaling operator localized there with scaling dimension $\Delta$, while $\HBCFT$ is the unperturbed boundary-CFT fixed-point Hamiltonian. The coupling $\lambda_t$ is controlled externally; $\alpha$ fixes its overall amplitude and $g(t)$ is a dimensionless protocol. In units with $\hbar=k_B=1$, $\alpha$ has energy dimension with scaling $1-\Delta$. The initial Gibbs state is $\pi_0=e^{-\beta\HBCFT}/Z_0$, where $\beta=(k_BT)^{-1}$, since $g(0)=0$. We impose boundary conditions $g(0)=0$, $g(\tau)=1$, and $|\alpha|\ll1$ so that the system may be described by linear response theory \cite{Kubo1957}. 

The boundary driven protocol does work on the system, which we treat as a stochastic variable $W$ with probability distribution $P(W)$ according to the standard two-time measurement construction. The statistical moments of work are determined via the moment generating function $G(\eta)= \ \mathbb{E}[e^{i\eta W}]$, which for the two-time measurement scheme becomes 
\begin{align}
    G(\eta)=\big\langle e^{i \eta H^H_\tau} e^{-i \eta H_0} \big\rangle_0
\end{align}
where $\langle.\rangle_0$ denotes the expectation with respect to equilibrium state $\pi_0$ and $H^H_t$ denotes the Heisenberg picture. Generalized linear response expresses every work cumulant as a quadratic functional of the protocol derivative and a relaxation kernel \cite{Guarnieri2024}. In the weak-driving regime, we focus on the second cumulant as the fluctuation cost to be constrained, which under linear response theory can be expressed as a positive quadratic functional of the protocol velocity:
\begin{equation}
\beta^2\sigma^2_{\scriptscriptstyle W}[g]=\alpha^2\int_0^\tau\dd t\int_0^\tau\dd t'\,
\dot g(t)\mathcal{C}(t-t',\beta)\dot g(t') .
\label{eq:Var}
\end{equation}
Here $\mathcal{C}(t,\beta)$ is the real-time equilibrium fluctuation kernel of the boundary operator, defined by
\begin{align}
    \mathcal{C}(t,\beta)=\frac{1}{2}\big\langle\{e^{i t \HBCFT}\delta \mathcal{O}_Be^{-i t \HBCFT},\delta \mathcal{O}_B\}\big\rangle_0
\end{align}
where we use $\mathcal{O}_B=\mathcal{O}_B(x=0)$ and  $\delta \mathcal{O} =\mathcal{O} -\langle \mathcal{O} \rangle_0$ for shorthand. For a boundary CFT, conformal invariance fixes the finite-temperature kernel to be \cite{AffleckLudwig1993,SM}
\begin{equation}
\mathcal{C}(t,\beta)=\beta^2C_{\Delta}\left[\frac{\pi/\beta}{|\sinh (\pi (t-i0^+)/\beta)|}\right]^{2\Delta}, 
\label{eq:O2}
\end{equation}
where $C_\Delta=\hbar\cos(\pi\Delta)/\Lambda^{2\Delta}$ in the normalization used here, and the $i0^+$ denotes an infinitesimal positive regulator at $t=0$. The formulas require $0<\Delta<1/2$ for an integrable short-time kernel; $\Delta=1/2$ is a separate local, marginal limit. 

Our goal is to now establish a speed limit bound relating the driving time $\tau$ and resulting work fluctuations. This becomes  a constrained quadratic variational problem since we want to minimise Eq.~\eqref{eq:Var} over all possible protocols $v(t)=\dot g(t)$ with constraint $\int_0^\tau v(t)\dd t=1$. This gives a Fredholm equation for the inverse-kernel profile $\zeta(t)=\hat K^{-1}1$, where $\hat K$ is an operator defined by kernel $\mathcal{C}(t,\beta)$. Using Eq.~\eqref{eq:O2}, the finite interval and the thermal scale can be disentangled by the conformal map $\mu(t)=\tanh[\pi (t-\tau/2\beta)]$. Since $\mu\in[-a,a]$, with $a=\tanh(\pi \tau/2\beta)$, the further rescaling $x=\mu/a$ maps the problem to $[-1,1]$. By introducing a new unknown function $u(x)$, the Fredholm equation becomes
\begin{equation}
\int_{-1}^1\frac{u(x)}{|y-x|^{2\Delta}}\dd x
=\lambda_\Delta(1-a^2y^2)^{-\Delta},
\label{eq:carleman-main}
\end{equation}
where the coefficient is $\lambda_\Delta=1/(\beta^2C_\Delta)$. (It is normalization-dependent, rather than scale-independent.) Thus the original inverse-kernel profile is recovered from $u$ through
\begin{equation}
\zeta(t,\beta)=\left(\frac{a}{\pi T}\right)^{2\Delta-1}
(1-a^2x^2)^{1-\Delta}u(x).
\label{eq:inverse-map-main}
\end{equation}
Equations~\eqref{eq:carleman-main} and \eqref{eq:inverse-map-main} show that the finite-temperature problem reduces to a compact inverse-kernel problem. The conformal map converts the boundary kernel into a Riesz-type kernel on a finite interval. Such weakly singular power-law kernels are standard objects in potential theory \cite{Landkof1972} and are closely related to fractional integration \cite{Samko1993}.

\begin{figure}[t]
\centering
{\includegraphics[width=1\linewidth]{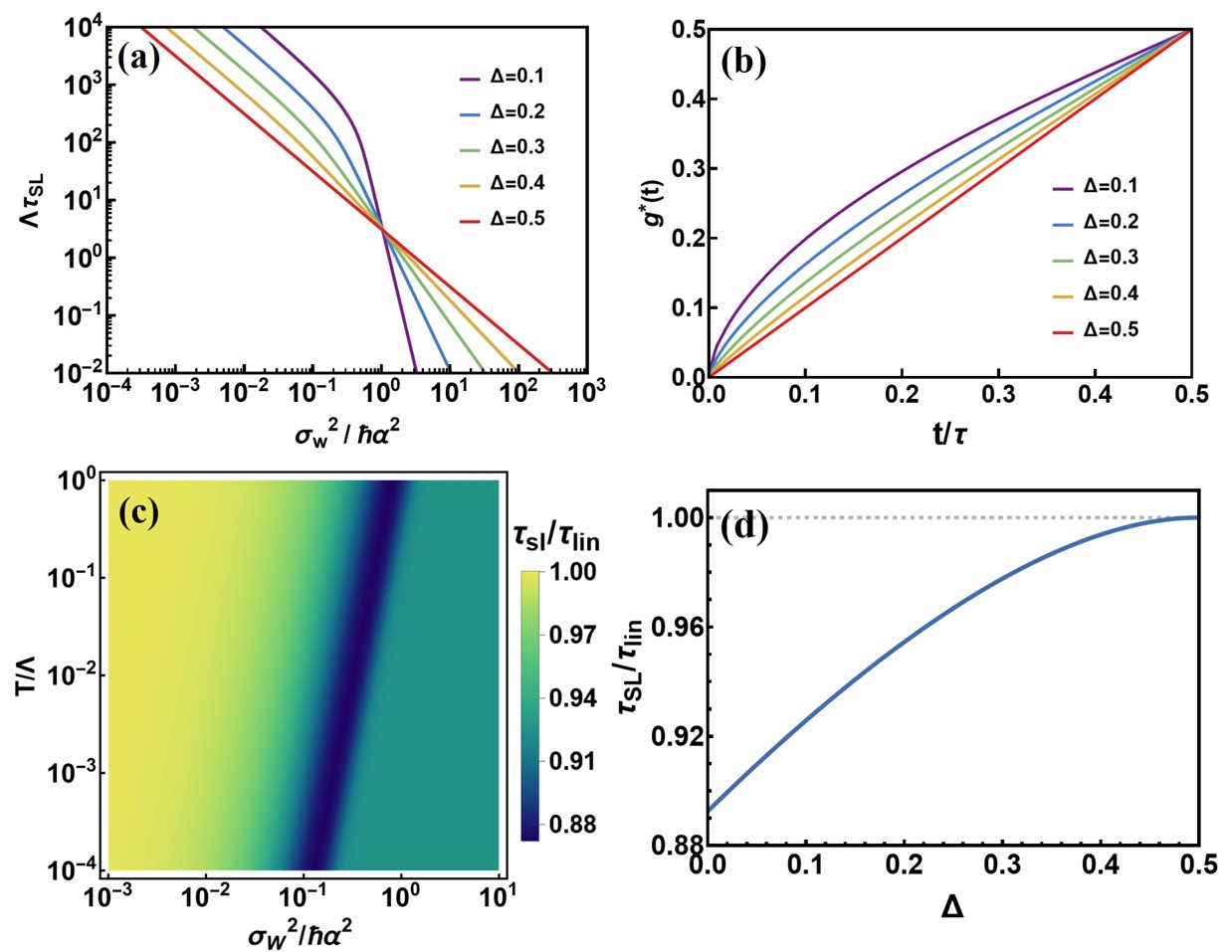}}
\caption{\label{fig:speedlimits}
Finite-temperature fluctuation speed limit and its saturating protocol. 
(a),(b) 
Speed limits and saturating protocols for $\Delta=0.1,0.2,0.3,0.4,$ and $0.5$ at $\Lambda\beta=100$. 
(c) Ratio $\tau_{\mathrm{SL}}/\tau_{\mathrm{lin}}$ as a function of temperature $T/\Lambda$ and reduced work fluctuation $\sigma^2_{\scriptscriptstyle W}/\hbar\alpha^2$ for $\Delta=0.1$. (d) Analytical Kibble--Zurek-regime ratio $\tau_{\mathrm{SL}}/\tau_{\mathrm{lin}}$ as a function of $\Delta$.}

\end{figure}

\textit{Finite temperature speed limit.}--- By using fractional calculus methods we are able to minimise the fluctuation over protocols. Since the work fluctuations are monotonically decreasing in time, we may invert the relation to turn this into a speed limit bound $\tau\geq\tau_{\mathrm{SL}}$ where \cite{SM}
\begin{equation}
    \tau_{\mathrm{SL}}=\frac{2}{\pi T}\operatorname{arctanh}
\left[\mathcal F_\Delta^{-1}\left(
\frac{\pi\hbar\alpha^{2}(\pi T)^{2\Delta}}
{\Lambda^{2\Delta}B\left(\frac12,\Delta+\frac12\right)\sigma^2_{\scriptscriptstyle W}}
\right)\right],
\label{eq:SL}
\end{equation}
where $\mathcal F_\Delta(z)=z^{2\Delta}{}_{2}F_{1}(1,\Delta;1+\Delta;z^{2})=\Delta B_{z^2}(\Delta,0)$, equivalently an incomplete-beta function, and $\mathcal F^{-1}$ is its inverse on $0<z^2<1$. The saturated protocol $g^*(t)$ can be obtained from Eq.~\eqref{eq:inverse-map-main} as 
\begin{align}
    g^*(t,\beta)=
\frac{\displaystyle\int_0^t \zeta(s,\beta)\,\mathrm{d}s}
     {\displaystyle\int_0^\tau \zeta(s,\beta)\,\mathrm{d}s}.
\end{align}

Equation~\eqref{eq:SL} is our central result and represents a tight speed limit for boundary CFTs at finite-temperature. It is useful to compare it with standard speed-limit bounds.  Quantum speed limits constrain the time required for state evolution in terms of instantaneous energy fluctuations and distinguishability \cite{MandelstamTamm1945,deffner2017quantum}.  Thermodynamic speed limits constrain the rate of non-equilibrium processes through entropy production or activity \cite{funo2019speed,van2021geometrical,Funo2025,Hasegawa2023,Yamauchi2025}.  By contrast, Eq.~\eqref{eq:SL} provides the minimum driving time relative to a given variance in non-equilibrium work along with an explicit protocol that saturates the bound. 

Figure~\ref{fig:speedlimits}(a) shows the speed limit. For $\Delta=1/2$, Eq.~\eqref{eq:C2CK} gives $\tau\propto[\sigma_{\scriptscriptstyle W}^{2}/\hbar\alpha^{2}]^{-1}$ throughout the temperature range considered, corresponding to a straight line of slope $-1$ on the log--log plot. In the upper-left region, where $\tau\gg\beta$, the adiabatic approximation applies, and all curves exhibit this inverse scaling, as described by Eq.~\eqref{eq:SLAL}. For $0<\Delta<1/2$, the curved portions around $\tau\sim\beta$ mark the finite-temperature crossover toward Kibble--Zurek scaling, with slope $-1/(2\Delta)$. We will discuss this regime in detail in the subsequent section. Figure~\ref{fig:speedlimits}(b) shows the saturating protocols for different scaling dimensions $\Delta$. Only the first half of each protocol is displayed, owing to symmetry about the midpoint of the drive. As $\Delta$ increases, the optimal protocol approaches a linear ramp, reflecting the fact that the critical kernel becomes local at $\Delta=1/2$:
\begin{equation}
\mathcal{C}^{\Delta=\frac{1}{2}}(t,\beta)=\frac{\pi\hbar\beta^2}{\Lambda}\delta(t).
\end{equation}
It follows that the inverse-kernel profile is the time-independent constant $\zeta(t)=\Lambda/\pi\beta^2\hbar$. Therefore the speed limit and optimal protocol are
\begin{equation}
  \tau_{\mathrm{SL}}^{\Delta=\frac{1}{2}}=\frac{\pi\hbar\alpha^2}{\Lambda\sigma^2_{\scriptscriptstyle W}},
  \quad
g_{\Delta=\frac{1}{2}}^*(t)=\frac{t}{\tau}.
\label{eq:C2CK}
\end{equation}
The $\Delta<1/2$ beta-function profile connects smoothly to this limit, becoming linear as $\Delta\to1/2$. In general the largest deviations from the  linear ramp occur at smaller $\Delta$, corresponding to an increase in non-local correlations in the fluctuation kernel.

For comparison, the linear protocol $g_{\operatorname{lin}}(t)=t/\tau$ gives
\begin{equation}
\beta^2 \sigma^2_{\scriptscriptstyle W}[g_{\operatorname{lin}}]
=\frac{2\alpha^2}{\tau^2}\int_0^\tau \mathrm{d}s\,(\tau-s)\mathcal{C}(s,\beta).
\end{equation}
Solving the optimized and linear relations at the same fluctuation tolerance $\sigma^2_{\scriptscriptstyle W}$ defines $\tau_{\mathrm{SL}}$ and $\tau_{\mathrm{lin}}$, respectively. Figure \ref{fig:speedlimits}(c) compares the optimised versus linear protocols. It shows $\tau_{\mathrm{SL}}/\tau_{\mathrm{lin}}$ as a function of temperature and work variance $\sigma^2_{\scriptscriptstyle W}/\hbar\alpha^2$ for $\Delta=0.1$. 

Values below unity quantify the gain from optimizing the protocol relative to a linear ramp.  We see that for small fluctuations and high temperatures, no improvement is found. This represents the adiabatic regime; indeed, by taking $\tau \gg\beta$ and hence $a\rightarrow1$ in Eq.~\eqref{eq:inverse-map-main}, the inverse-kernel profile becomes constant to leading order. Normalization then gives the fluctuation-optimal protocol as a linear ramp $g^*(t)\simeq t/\tau$.

\textit{Kibble-Zurek regime.}--- In Figure \ref{fig:speedlimits} (c) we observe that there is a diagonal crossover along the curve $T\sim \sigma_{\scriptscriptstyle W}^{1/\Delta}$, equivalently $\mathrm{d}\log T/\mathrm{d}\log\sigma_{\scriptscriptstyle W}=1/\Delta$. This represents a cross-over into the universal finite-time critical regime  in which the driving time lies between the microscopic and thermal timescales, $1/\Lambda\ll\tau \ll\beta$. This is known as the Kibble-Zurek regime in which the finite-temperature conformal kernel reduces to its scale-free zero-temperature form.

The speed limit and its optimsied protocol can here be obtained analytically. Expanding Eq.~\eqref{eq:SL} for small $T\tau$ gives the minimum driving time in the KZ regime:
\begin{equation}
\tau^{\KZ}_{\mathrm{SL}}
=\left(\frac{\pi\hbar\alpha^{2}}{\Lambda^{2\Delta}B\left(\Delta+\frac{1}{2},\Delta+\frac{1}{2}\right)\sigma^2_{\scriptscriptstyle W}}\right)^{1/2\Delta}.\label{eq:SLKZ}
\end{equation}
This represents the minimum driving timescale whenever the work fluctuations are within the interval
\begin{align}
    \bigg(\frac{T}{\Lambda}\bigg)^{\Delta}\ll \frac{\sigma_W}{\sqrt{\hbar}\alpha}\ll 1.
    \label{eq:KZforVar}
\end{align}
Returning also to Figure~\ref{fig:speedlimits}(a) we can see that the straight-line portions in the lower-right region follow the $\tau T\ll1$ asymptote of the theoretical prediction with slope $-1/(2\Delta)$. Although they continue the same power law, they fall outside the controlled KZ regime when $\Lambda\tau\gg1$ is no longer satisfied.

The optimal driving protocol can also be obtained explicitly. The source in Eq.~\eqref{eq:carleman-main} is constant to leading order, and the inverse Riesz kernel produces an endpoint-enhanced inverse-kernel profile. After normalization, the corresponding  optimal protocol is
\begin{equation}
g_{\KZ}^*(t)
=I_{t/\tau}\!\left(\Delta+\frac12,\Delta+\frac12\right),
\label{eq:kz-var-protocol}
\end{equation}
where $I_x(a,b)$ is the regularized incomplete beta function.
In this limit, Eq. \eqref{eq:SLKZ} is the temperature-independent critical limit of the universal speed limit. It gives a saturable lower bound on the protocol duration, with $\tau_{\mathrm{SL}}\propto\sigma_{\scriptscriptstyle W}^{-1/\Delta}$. The dependence on $\Delta$ is universal and enters both the bound and the saturating protocol. Equation~\eqref{eq:kz-var-protocol} shows that the optimal profile is a beta-function ramp: for $\Delta<1/2$ its velocity is enhanced near the endpoints, while in the limit $\Delta\to1/2$ it reduces smoothly to the uniform ramp.

\begin{figure*}[t]
\centering
\IfFileExists{Fig2are.jpg}{\includegraphics[scale=0.32]{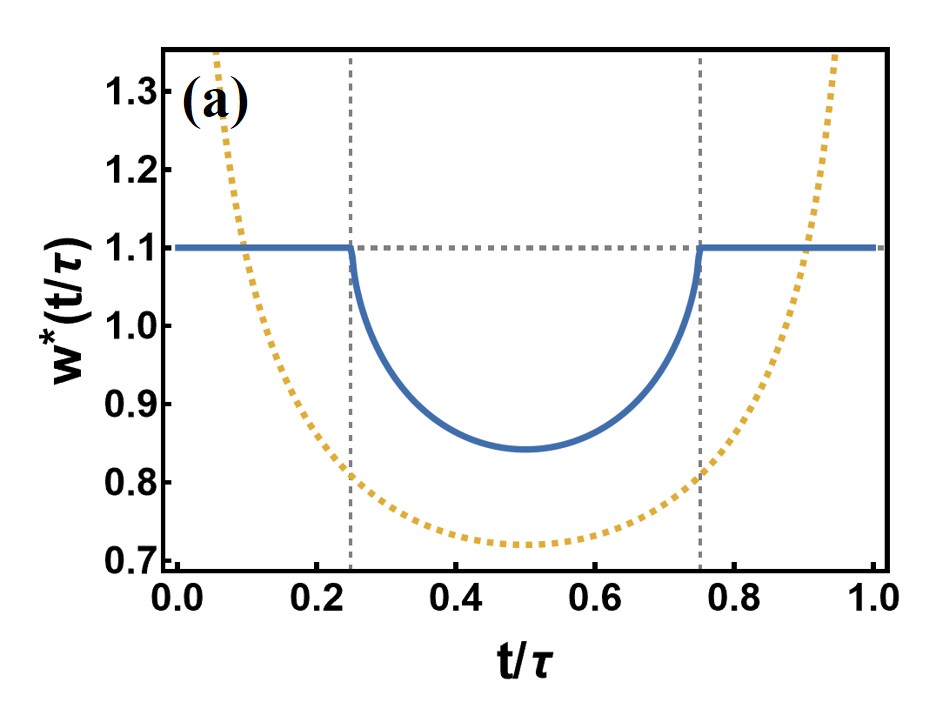}}{\fbox{\parbox[c][0.16\textheight][c]{0.8\linewidth}{\centering Missing figure: Fig2a.jpg}}}
\IfFileExists{Fig2bre.jpg}{\includegraphics[scale=0.32]{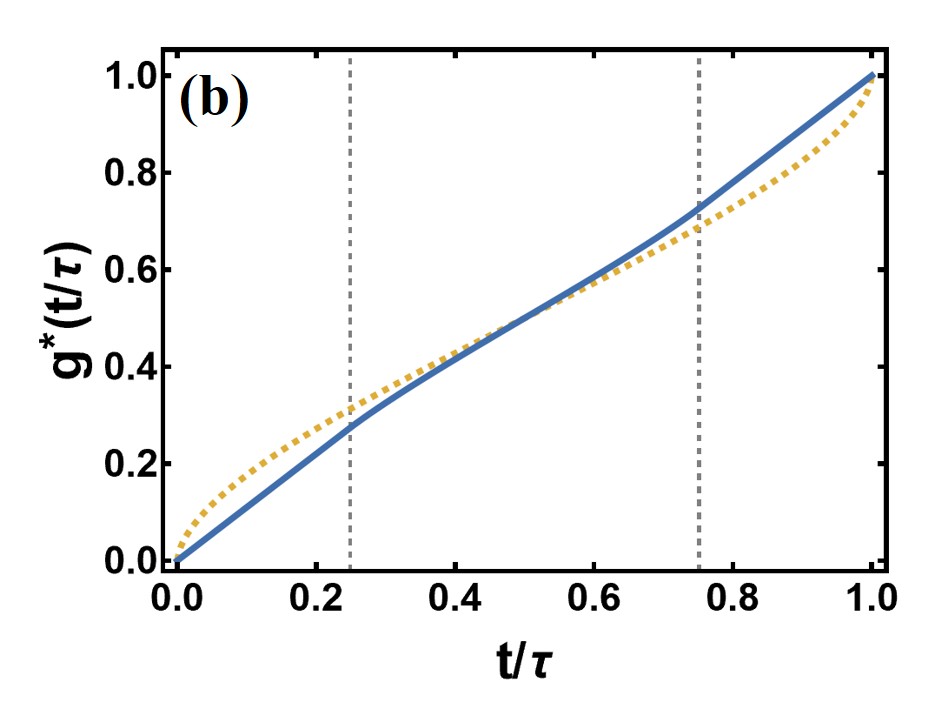}}{\fbox{\parbox[c][0.16\textheight][c]{0.8\linewidth}{\centering Missing figure: Fig2b.jpg}}}
\IfFileExists{Fig2cre.jpg}{\includegraphics[scale=0.32]{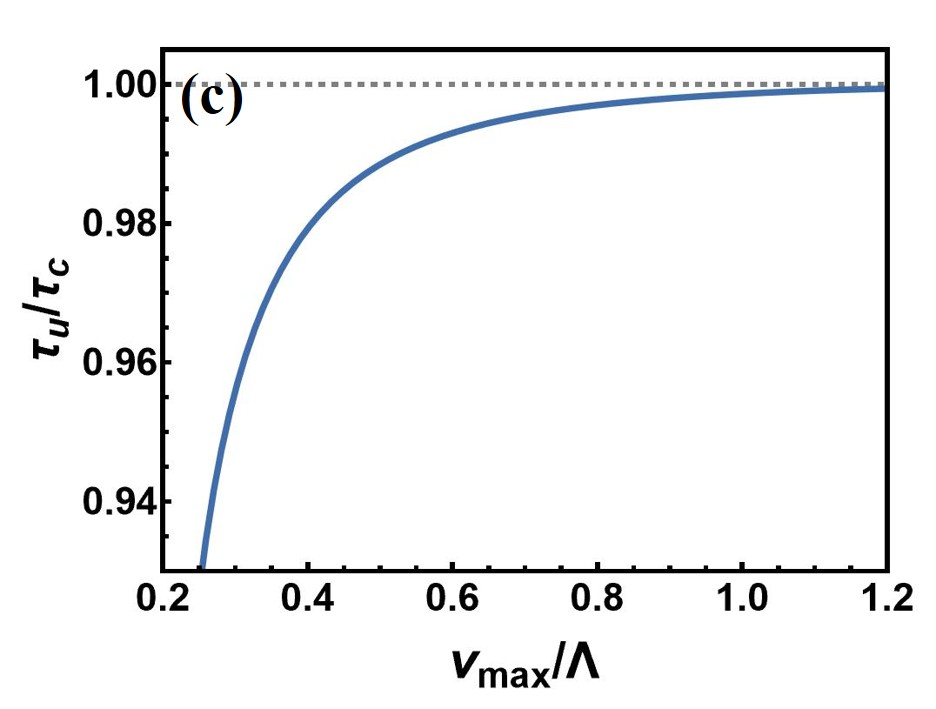}}{\fbox{\parbox[c][0.16\textheight][c]{0.8\linewidth}{\centering Missing figure: Fig2c.jpg}}}
\caption{\label{fig:CLPTs}
Velocity-constrained optimal control in the Kibble--Zurek regime.
(a) Optimal dimensionless velocity $w^*(t/\tau)=\tau v^*(t)$.
(b) Corresponding optimal protocol $g^*(t/\tau)$.
In (a) and (b), the yellow dashed curves represent the unconstrained solutions, while the blue solid curves represent the velocity-constrained solutions with $w_{\max}=1.1$.
(c) Ratio $\tau_u/\tau_c$ between the unconstrained and velocity-constrained speed-limit times as a function of $v_{\max}/\Lambda$ for $\Delta=0.1$.
}
\end{figure*}
To compare this with a linear ramp one can directly calculate the time as a function of work fluctuations,
\begin{equation}
\tau^{\KZ}_{\mathrm{lin}}
=
\left[
\frac{\pi\hbar\alpha^2}
{2(1-\Delta)\Lambda^{2\Delta}
B\left(\Delta+\frac12,\frac32-\Delta\right)\sigma^2_{\scriptscriptstyle W}}
\right]^{1/(2\Delta)} ,
\label{eq:linKZ}
\end{equation}
Combining Eqs.~\eqref{eq:SLKZ} and \eqref{eq:linKZ}, the work variance cancels and one obtains a constant ratio
\begin{equation}
\frac{\tau^{\KZ}_{\mathrm{SL}}}
{\tau^{\KZ}_{\mathrm{lin}}}
=
\left[
2(1-\Delta)
\frac{
B\left(\Delta+\frac12,\frac32-\Delta\right)
}{
B\left(\Delta+\frac12,\Delta+\frac12\right)
}
\right]^{1/(2\Delta)}.
\label{eq:SL-lin-ratio}
\end{equation}
Fig. \ref{fig:speedlimits}(d) shows this analytical result, where the ratio approaches unity as $\Delta\to1/2$, consistent with the disappearance of the optimization gain near the local-kernel threshold. Finally, we observe that the largest improvement of optimal protocols is found within the cross-over regime, as indicated by the blue region in Fig. \ref{fig:speedlimits}(c). 

\textit{Velocity constraints.}--- Next we observe that for the optimal solution Eq. ~\eqref{eq:kz-var-protocol}, the velocity of the coupling diverges at the endpoints of the protocol. For practical implementations this motivates imposing a finite-bandwidth constraint on the protocol velocity, 
\begin{align}
    0\leq v(t)=\dot g(t)\leq v_{\max},
\end{align}
and investigating how robust the speed limit is. Since the fluctuation cost is a positive quadratic functional and the constraints are linear, the constrained optimum is characterized by the Karush--Kuhn--Tucker conditions
\cite{Karush1939,KuhnTucker1951,BoydVandenberghe2004}. 
For the solutions considered here the lower bound is inactive, and the optimal velocity satisfies
\begin{equation}
\int_0^\tau dt'\,
\frac{v^*(t')}{|t-t'|^{2\Delta}}
=
\lambda-\xi(t),
\label{eq:kkt-velocity}
\end{equation}
with $\xi(t)\geq0$,  $\xi(t)[v_{\max}-v^*(t)]=0$, and the normalization $\int_0^\tau dt\,v^*(t)=1$. Here $\lambda$ fixes the total displacement, while $\xi(t)$ is nonzero only where the velocity constraint is active. Thus the constrained optimum consists of saturated arcs with $v^*(t)=v_{\max}$, matched to a free segment where $\xi(t)=0$ and the inverse-kernel equation is recovered.

 In Fig.~\ref{fig:CLPTs}(a), the unconstrained Kibble--Zurek optimum enhances the velocity near the endpoints and exceeds the imposed bound. The constrained solution therefore clips these endpoint regions into saturated arcs, with switching points marked by the vertical dashed lines. 
Figure~\ref{fig:CLPTs}(b) shows the corresponding protocol: the saturated arcs become linear pieces, while the middle part remains a nontrivial free solution. This active-set restructuring of the optimal protocol is reminiscent of the
structural changes in optimal solutions that underlie control-landscape phase transitions \cite{Beato2025}. Figure~\ref{fig:CLPTs}(c) quantifies the time penalty of the velocity bound through $\tau_u/\tau_c$, where $\tau_u$ and $\tau_c$ are the unconstrained and constrained speed-limit times. 
Since the constraint can only increase the minimum duration, $\tau_u/\tau_c\leq1$. 
As $v_{\max}/\Lambda$ increases, the constraint becomes inactive and the ratio approaches unity.

\textit{Experimental outlook.---} Our speed limit applies to a wide range of experimentally relevant systems that can be described by the finite-temperature kernel~\eqref{eq:O2} in the weak driving regime. As we have seen the speed limit has greater impact for a low scaling dimension $\Delta$, thus one should consider physical settings where this low scaling can be realised. Fractional quantum Hall point contacts provide one realization, where a gate is used to control quasiparticle tunnelling between edge states \cite{BertinJohannet2023}. For a Laughlin state with filling fraction $\nu$, the tunnelling operator has scaling dimension $\Delta=\nu$. Bulk Laughlin states with $\nu=1/5$ (thus $\Delta=0.2$) have been  experimentally observed \cite{Roddaro2004}. Multichannel charge-Kondo circuits
are another realization, with dimensionality set by the number of leads $M$ via $\Delta=2/(M+2)$. While two- and three-channel criticality has been observed experimentally \cite{Iftikhar2015,Iftikhar2018}, larger numbers of leads represent a future engineering challenge. A particularly promising platform for probing the low $\Delta$ regime are superconducting
boundary-sine--Gordon circuits controlled using flux-tunable Josephson junctions \cite{Leger2023}.  Here the scaling dimension of the 
Josephson operator is set by the ratio of environment impedance to superconducting resistance, $\Delta=Z/R_Q$, making the scaling dimension
continuously engineerable. Fast
time-dependent SQUID modulation is already experimentally established
\cite{Leger2019,Kuzmin2025}, and so these circuits provide a
natural platform for testing the speed limit.

\textit{Conclusion.---} We have derived a finite temperature and saturable work fluctuation speed limit for any weakly-driven quantum system described by a boundary CFT. Our results identify quantum work fluctuations as a key resource for non-equilibrium quantum control in critical regimes, and demonstrate that the scaling dimension can be used as an operational measure of the precision--time control cost. While we have highlighted a number of experimentally relevant
critical many-body systems where this speed limit is applicable, it would also be
interesting to apply these methods to holographic settings. Recent work has established a holographic dual of the quantum work distribution in AdS/CFT \cite{Takeda2026}, raising the possibility that fluctuation speed limits such as those derived here may admit a direct gravitational interpretation.

\textit{Acknowledgments.---} S.X. acknowledges support from The University of Manchester under the Dean’s Doctoral Scholarship.  HJDM acknowledges funding from a Royal Society Research Fellowship (URF/R1/231394). We thank Andrew Mitchell and Pete Millington for helpful discussions.

\bibliographystyle{apsrev4-2}
\bibliography{refs}

@article{cerisola2017using,
  title={Using a quantum work meter to test non-equilibrium fluctuation theorems},
  author={Cerisola, Federico and Margalit, Yair and Machluf, Shimon and Roncaglia, Augusto J and Paz, Juan Pablo and Folman, Ron},
  journal={Nature communications},
  volume={8},
  number={1},
  pages={1241},
  year={2017},
  publisher={Nature Publishing Group UK London},
  doi = {10.1038/s41467-017-01308-7},
  url = {https://doi.org/10.1038/s41467-017-01308-7}
}

@article{funo2017universal,
  title={Universal work fluctuations during shortcuts to adiabaticity by counterdiabatic driving},
  author={Funo, Ken and Zhang, Jing-Ning and Chatou, Cyril and Kim, Kihwan and Ueda, Masahito and Del Campo, Adolfo},
  journal={Physical Review Letters},
  volume={118},
  number={10},
  pages={100602},
  year={2017},
  publisher={APS},
  doi = {10.1103/PhysRevLett.118.100602},
  url = {https://doi.org/10.1103/PhysRevLett.118.100602}
}

@article{hasegawa2021thermodynamic,
  title={Thermodynamic uncertainty relation for general open quantum systems},
  author={Hasegawa, Yoshihiko},
  journal={Physical Review Letters},
  volume={126},
  number={1},
  pages={010602},
  year={2021},
  publisher={APS},
  doi = {10.1103/PhysRevLett.126.010602},
  url = {https://doi.org/10.1103/PhysRevLett.126.010602}
}

@article{miller2021thermodynamic,
  title={Thermodynamic uncertainty relation in slowly driven quantum heat engines},
  author={Miller, Harry JD and Mohammady, M Hamed and Perarnau-Llobet, Mart{\'\i} and Guarnieri, Giacomo},
  journal={Physical review letters},
  volume={126},
  number={21},
  pages={210603},
  year={2021},
  publisher={APS},
  doi = {10.1103/PhysRevLett.126.210603},
  url = {https://doi.org/10.1103/PhysRevLett.126.210603}
}

@article{van2021geometrical,
  title={Geometrical bounds of the irreversibility in Markovian systems},
  author={Van Vu, Tan and Hasegawa, Yoshihiko},
  journal={Physical Review Letters},
  volume={126},
  number={1},
  pages={010601},
  year={2021},
  publisher={APS},
  doi = {10.1103/PhysRevLett.126.010601},
  url = {https://doi.org/10.1103/PhysRevLett.126.010601}
}

@article{funo2019speed,
  title={Speed limit for open quantum systems},
  author={Funo, Ken and Shiraishi, Naoto and Saito, Keiji},
  journal={New Journal of Physics},
  volume={21},
  number={1},
  pages={013006},
  year={2019},
  publisher={IOP Publishing},
  doi = {10.1088/1367-2630/aaf9f5},
  url = {https://doi.org/10.1088/1367-2630/aaf9f5}
}

@article{van2023thermodynamic,
  title={Thermodynamic unification of optimal transport: Thermodynamic uncertainty relation, minimum dissipation, and thermodynamic speed limits},
  author={Van Vu, Tan and Saito, Keiji},
  journal={Physical Review X},
  volume={13},
  number={1},
  pages={011013},
  year={2023},
  publisher={APS},
  doi = {10.1103/PhysRevX.13.011013},
  url = {https://doi.org/10.1103/PhysRevX.13.011013}
}

@article{del2013quantum,
  title={Quantum speed limits in open system dynamics},
  author={del Campo, Adolfo and Egusquiza, Inigo L and Plenio, Martin B and Huelga, Susana F},
  journal={Physical review letters},
  volume={110},
  number={5},
  pages={050403},
  year={2013},
  publisher={APS},
  doi = {10.1103/PhysRevLett.110.050403},
  url = {https://doi.org/10.1103/PhysRevLett.110.050403}
}

@article{deffner2017quantum,
  title={Quantum speed limits: from Heisenberg’s uncertainty principle to optimal quantum control},
  author={Deffner, Sebastian and Campbell, Steve},
  journal={Journal of Physics A: Mathematical and Theoretical},
  volume={50},
  number={45},
  pages={453001},
  year={2017},
  publisher={IOP Publishing},
  doi = {10.1088/1751-8121/aa86c6},
  url = {https://doi.org/10.1088/1751-8121/aa86c6}
}

@incollection{funo2019quantum,
  title={Quantum fluctuation theorems},
  author={Funo, Ken and Ueda, Masahito and Sagawa, Takahiro},
  booktitle={Thermodynamics in the Quantum Regime: Fundamental Aspects and New Directions},
  pages={249--273},
  year={2019},
  publisher={Springer},
  doi = {10.1007/978-3-319-99046-0_10},
  url = {https://doi.org/10.1007/978-3-319-99046-0_10}
}

@article{batalhao2014experimental,
  title={Experimental reconstruction of work distribution and study of fluctuation relations in a closed quantum system},
  author={Batalh{\~a}o, Tiago B and Souza, Alexandre M and Mazzola, Laura and Auccaise, Ruben and Sarthour, Roberto S and Oliveira, Ivan S and Goold, John and De Chiara, Gabriele and Paternostro, Mauro and Serra, Roberto M},
  journal={Physical review letters},
  volume={113},
  number={14},
  pages={140601},
  year={2014},
  publisher={APS},
  doi = {10.1103/PhysRevLett.113.140601},
  url = {https://doi.org/10.1103/PhysRevLett.113.140601}
}

@article{campisi2011colloquium,
  title={Colloquium: Quantum fluctuation relations: Foundations and applications},
  author={Campisi, Michele and H{\"a}nggi, Peter and Talkner, Peter},
  journal={Reviews of Modern Physics},
  volume={83},
  number={3},
  pages={771--791},
  year={2011},
  publisher={APS},
  doi = {10.1103/RevModPhys.83.771},
  url = {https://doi.org/10.1103/RevModPhys.83.771}
}

@article{naze2022kibble,
  title={Kibble--Zurek scaling from linear response theory},
  author={Naz{\'e}, Pierre and Bonan{\c{c}}a, Marcus VS and Deffner, Sebastian},
  journal={Entropy},
  volume={24},
  number={5},
  pages={666},
  year={2022},
  publisher={MDPI},
  doi = {10.3390/e24050666},
  url = {https://doi.org/10.3390/e24050666}
}

@article{naze2022optimal,
  title={Optimal finite-time processes in weakly driven overdamped Brownian motion},
  author={Naz{\'e}, Pierre and Deffner, Sebastian and Bonan{\c{c}}a, Marcus VS},
  journal={Journal of Physics Communications},
  volume={6},
  number={8},
  pages={083001},
  year={2022},
  publisher={IOP Publishing},
  doi = {10.1088/2399-6528/ac871d},
  url = {https://doi.org/10.1088/2399-6528/ac871d}
}

@article{bonancca2014optimal,
  title={Optimal driving of isothermal processes close to equilibrium},
  author={Bonan{\c{c}}a, Marcus VS and Deffner, Sebastian},
  journal={The Journal of chemical physics},
  volume={140},
  number={24},
  year={2014},
  publisher={AIP publishing},
  doi = {10.1063/1.4885277},
  url = {https://doi.org/10.1063/1.4885277}
}

@article{doria2011optimal,
  title={Optimal control technique for many-body quantum dynamics},
  author={Doria, Patrick and Calarco, Tommaso and Montangero, Simone},
  journal={Physical review letters},
  volume={106},
  number={19},
  pages={190501},
  year={2011},
  publisher={APS},
  doi = {10.1103/PhysRevLett.106.190501},
  url = {https://doi.org/10.1103/PhysRevLett.106.190501}
}

@article{b2020universal,
  title={Universal finite-time thermodynamics of many-body quantum machines from Kibble-Zurek scaling},
  author={B. S, Revathy and Mukherjee, Victor and Divakaran, Uma and del Campo, Adolfo},
  journal={Physical Review Research},
  volume={2},
  number={4},
  pages={043247},
  year={2020},
  publisher={APS},
  doi = {10.1103/PhysRevResearch.2.043247},
  url = {https://doi.org/10.1103/PhysRevResearch.2.043247}
}

@article{heyl2017quenching,
  title={Quenching a quantum critical state by the order parameter: Dynamical quantum phase transitions and quantum speed limits},
  author={Heyl, Markus},
  journal={Physical Review B},
  volume={95},
  number={6},
  pages={060504},
  year={2017},
  publisher={APS},
  doi = {10.1103/PhysRevB.95.060504},
  url = {https://doi.org/10.1103/PhysRevB.95.060504}
}

@article{del2012assisted,
  title={Assisted finite-rate adiabatic passage across a quantum critical point: Exact solution for the quantum Ising model},
  author={del Campo, Adolfo and Rams, Marek M and Zurek, Wojciech H},
  journal={Physical review letters},
  volume={109},
  number={11},
  pages={115703},
  year={2012},
  publisher={APS},
  doi = {10.1103/PhysRevLett.109.115703},
  url = {https://doi.org/10.1103/PhysRevLett.109.115703}
}

@article{brandner2020thermodynamic,
  title={Thermodynamic geometry of microscopic heat engines},
  author={Brandner, Kay and Saito, Keiji},
  journal={Physical review letters},
  volume={124},
  number={4},
  pages={040602},
  year={2020},
  publisher={APS},
  doi = {10.1103/PhysRevLett.124.040602},
  url = {https://doi.org/10.1103/PhysRevLett.124.040602}
}

@article{erdman2023pareto,
  title={Pareto-optimal cycles for power, efficiency and fluctuations of quantum heat engines using reinforcement learning},
  author={Erdman, Paolo A and Rolandi, Alberto and Abiuso, Paolo and Perarnau-Llobet, Mart{\'\i} and No{\'e}, Frank},
  journal={Physical Review Research},
  volume={5},
  number={2},
  pages={L022017},
  year={2023},
  publisher={APS},
  doi = {10.1103/PhysRevResearch.5.L022017},
  url = {https://doi.org/10.1103/PhysRevResearch.5.L022017}
}

@article{cavina2018optimal,
  title={Optimal thermodynamic control in open quantum systems},
  author={Cavina, Vasco and Mari, Andrea and Carlini, Alberto and Giovannetti, Vittorio},
  journal={Physical Review A},
  volume={98},
  number={1},
  pages={012139},
  year={2018},
  publisher={APS},
  doi = {10.1103/PhysRevA.98.012139},
  url = {https://doi.org/10.1103/PhysRevA.98.012139}
}

@article{campbell2017trade,
  title={Trade-off between speed and cost in shortcuts to adiabaticity},
  author={Campbell, Steve and Deffner, Sebastian},
  journal={Physical review letters},
  volume={118},
  number={10},
  pages={100601},
  year={2017},
  publisher={APS},
  doi = {10.1103/PhysRevLett.118.100601},
  url = {https://doi.org/10.1103/PhysRevLett.118.100601}
}

@article{guery2019shortcuts,
  title={Shortcuts to adiabaticity: Concepts, methods, and applications},
  author={Gu{\'e}ry-Odelin, David and Ruschhaupt, Andreas and Kiely, Anthony and Torrontegui, Erik and Mart{\'\i}nez-Garaot, Sofia and Muga, Juan Gonzalo},
  journal={Reviews of Modern Physics},
  volume={91},
  number={4},
  pages={045001},
  year={2019},
  publisher={APS},
  doi = {10.1103/RevModPhys.91.045001},
  url = {https://doi.org/10.1103/RevModPhys.91.045001}
}

@article{abiuso2020geometric,
  title={Geometric optimisation of quantum thermodynamic processes},
  author={Abiuso, Paolo and Miller, Harry JD and Perarnau-Llobet, Mart{\'\i} and Scandi, Matteo},
  journal={Entropy},
  volume={22},
  number={10},
  pages={1076},
  year={2020},
  publisher={MDPI},
  doi = {10.3390/e22101076},
  url = {https://doi.org/10.3390/e22101076}
}

@article{acconcia2015shortcuts,
  title={Shortcuts to adiabaticity from linear response theory},
  author={Acconcia, Thiago V and Bonan{\c{c}}a, Marcus VS and Deffner, Sebastian},
  journal={Physical Review E},
  volume={92},
  number={4},
  pages={042148},
  year={2015},
  publisher={APS},
  doi = {10.1103/PhysRevE.92.042148},
  url = {https://doi.org/10.1103/PhysRevE.92.042148}
}

@article{Leger2019,
  author  = {L{\'e}ger, S{\'e}bastien and Puertas-Mart{\'i}nez, Javier and
             Bharadwaj, Karthik and Dassonneville, R{\'e}my and
             Delaforce, Jovian and Foroughi, Farshad and Milchakov, Vladimir
             and Planat, Luca and Buisson, Olivier and Naud, C{\'e}cile and
             Hasch-Guichard, Wiebke and Florens, Serge and Snyman, Izak
             and Roch, Nicolas},
  title   = {Observation of quantum many-body effects due to zero point
             fluctuations in superconducting circuits},
  journal = {Nature Communications},
  volume  = {10},
  pages   = {5259},
  year    = {2019},
  doi     = {10.1038/s41467-019-13199-x},
  url = {https://doi.org/10.1038/s41467-019-13199-x}
}

@article{Leger2023,
  author  = {L{\'e}ger, S{\'e}bastien and S{\'e}pulcre, Th{\'e}o and
             Fraudet, Dorian and Buisson, Olivier and Naud, C{\'e}cile and
             Hasch-Guichard, Wiebke and Florens, Serge and Snyman, Izak and
             Basko, Denis M. and Roch, Nicolas},
  title   = {Revealing the finite-frequency response of a bosonic quantum impurity},
  journal = {SciPost Physics},
  volume  = {14},
  pages   = {130},
  year    = {2023},
  doi     = {10.21468/SciPostPhys.14.5.130},
  url = {https://doi.org/10.21468/SciPostPhys.14.5.130}
}

@article{Kuzmin2025,
  author  = {Kuzmin, Roman and Mehta, Nitish and Grabon, Nicholas and
             Mencia, Raymond A. and Burshtein, Amir and Goldstein, Moshe
             and Manucharyan, Vladimir E.},
  title   = {Observation of the {Schmid--Bulgadaev} dissipative quantum
             phase transition},
  journal = {Nature Physics},
  volume  = {21},
  pages   = {132--136},
  year    = {2025},
  doi     = {10.1038/s41567-024-02695-7},
  url = {https://doi.org/10.1038/s41567-024-02695-7}
}

@article{Roddaro2004,
  author  = {Roddaro, Stefano and Pellegrini, Vittorio and Beltram, Fabio
             and Biasiol, Giorgio and Sorba, Lucia},
  title   = {Interedge Strong-to-Weak Scattering Evolution at a
             Constriction in the Fractional Quantum Hall Regime},
  journal = {Phys. Rev. Lett.},
  volume  = {93},
  pages   = {046801},
  year    = {2004},
  doi     = {10.1103/PhysRevLett.93.046801},
  url = {https://doi.org/10.1103/PhysRevLett.93.046801}
}

@book{Landkof1972,
  author    = {Landkof, N. S.},
  title     = {Foundations of Modern Potential Theory},
  series    = {Grundlehren der mathematischen Wissenschaften},
  volume    = {180},
  publisher = {Springer-Verlag},
  year      = {1972},
  doi = {10.1007/978-3-642-65183-0},
  url = {https://doi.org/10.1007/978-3-642-65183-0}
}

@book{Samko1993,
  author    = {Samko, Stefan G. and Kilbas, Anatoly A. and Marichev, Oleg I.},
  title     = {Fractional Integrals and Derivatives: Theory and Applications},
  publisher = {Gordon and Breach Science Publishers},
  year      = {1993},
  url = {https://ci.nii.ac.jp/ncid/BA21828325}
}

@article{Takeda2026,
  author  = {Takeda, Daichi},
  title   = {Work distribution and fluctuation theorem in AdS/CFT},
  journal = {Phys. Rev. D},
  volume  = {113},
  pages   = {106029},
  year    = {2026},
  doi     = {10.1103/6hw6-t8l7},
  url = {https://doi.org/10.1103/6hw6-t8l7}
}

@article{BertinJohannet2023,
  author  = {Bertin-Johannet, Bruno and Raymond, Laurent and Rech, J{\'e}r{\^o}me
             and Jonckheere, Thibaut and Gr{\'e}maud, Beno{\^i}t
             and Glattli, D. Christian and Martin, Thierry},
  title   = {Photoassisted current in the fractional quantum Hall effect
             as a probe of the quasiparticle operator scaling dimension},
  journal = {Physical Review B},
  volume  = {108},
  pages   = {165425},
  year    = {2023},
  doi     = {10.1103/PhysRevB.108.165425},
  url = {https://doi.org/10.1103/PhysRevB.108.165425}
}

@article{Vidunas_2011,
   title={A generalization of Clausen’s identity},
   volume={26},
   ISSN={1572-9303},
   url={http://dx.doi.org/10.1007/s11139-011-9314-1},
   DOI={10.1007/s11139-011-9314-1},
   number={1},
   journal={The Ramanujan Journal},
   publisher={Springer Science and Business Media LLC},
   author={Vidunas, Raimundas},
   year={2011},
   month=Aug, pages={133–146} }

@article{Kibble1976,
  author = {Kibble, T. W. B.},
  title = {Topology of Cosmic Domains and Strings},
  journal = {J. Phys. A: Math. Gen.},
  volume = {9},
  pages = {1387--1398},
  year = {1976},
  doi = {10.1088/0305-4470/9/8/029},
  url = {https://doi.org/10.1088/0305-4470/9/8/029}
}

@article{Zurek1985,
  author = {Zurek, W. H.},
  title = {Cosmological Experiments in Superfluid Helium?},
  journal = {Nature},
  volume = {317},
  pages = {505--508},
  year = {1985},
  doi = {10.1038/317505a0},
  url = {https://doi.org/10.1038/317505a0}
}

@article{Dziarmaga2005,
  author = {Dziarmaga, Jacek},
  title = {Dynamics of a Quantum Phase Transition: Exact Solution of the Quantum Ising Model},
  journal = {Phys. Rev. Lett.},
  volume = {95},
  pages = {245701},
  year = {2005},
  doi = {10.1103/PhysRevLett.95.245701},
  url = {https://doi.org/10.1103/PhysRevLett.95.245701}
}

@article{Polkovnikov2005,
  author = {Polkovnikov, Anatoli},
  title = {Universal Adiabatic Dynamics in the Vicinity of a Quantum Critical Point},
  journal = {Phys. Rev. B},
  volume = {72},
  pages = {161201(R)},
  year = {2005},
  doi = {10.1103/PhysRevB.72.161201},
  url = {https://doi.org/10.1103/PhysRevB.72.161201}
}

@article{Talkner2007,
  author = {Talkner, Peter and Lutz, Eric and H{\"a}nggi, Peter},
  title = {Fluctuation Theorems: Work Is Not an Observable},
  journal = {Phys. Rev. E},
  volume = {75},
  pages = {050102(R)},
  year = {2007},
  doi = {10.1103/PhysRevE.75.050102},
  url = {https://doi.org/10.1103/PhysRevE.75.050102}
}

@article{Fei2020,
  author = {Fei, Zongping and Freitas, Nahuel and Cavina, Vasco and Quan, H. T. and Esposito, Massimiliano},
  title = {Work Statistics across a Quantum Phase Transition},
  journal = {Phys. Rev. Lett.},
  volume = {124},
  pages = {170603},
  year = {2020},
  doi = {10.1103/PhysRevLett.124.170603},
  url = {https://doi.org/10.1103/PhysRevLett.124.170603}
}

@article{Guarnieri2024,
  author = {Guarnieri, Giacomo and Eisert, Jens and Miller, Harry J. D.},
  title = {Generalized Linear Response Theory for the Full Quantum Work Statistics},
  journal = {Phys. Rev. Lett.},
  volume = {133},
  pages = {070405},
  year = {2024},
  doi = {10.1103/PhysRevLett.133.070405},
  url = {https://doi.org/10.1103/PhysRevLett.133.070405}
}

@article{Ma2025,
  author = {Ma, Zhanyu and Mitchell, Andrew K. and Sela, Eran},
  title = {Quantum Work Statistics across a Critical Point: Full Crossover from Sudden Quench to the Adiabatic Limit},
  journal = {Phys. Rev. Lett.},
  volume = {135},
  pages = {130402},
  year = {2025},
  doi = {10.1103/vn83-mt2v},
  url = {https://doi.org/10.1103/vn83-mt2v}
}

@article{Kubo1957,
  author = {Kubo, Ryogo},
  title = {Statistical-Mechanical Theory of Irreversible Processes. I. General Theory and Simple Applications to Magnetic and Conduction Problems},
  journal = {J. Phys. Soc. Jpn.},
  volume = {12},
  pages = {570--586},
  year = {1957},
  doi = {10.1143/JPSJ.12.570},
  url = {https://doi.org/10.1143/JPSJ.12.570}
}

@article{Cardy1989,
  author = {Cardy, John L.},
  title = {Boundary Conditions, Fusion Rules and the Verlinde Formula},
  journal = {Nucl. Phys. B},
  volume = {324},
  pages = {581--596},
  year = {1989},
  doi = {10.1016/0550-3213(89)90521-X},
  url = {https://doi.org/10.1016/0550-3213(89)90521-X}
}

@article{AffleckLudwig1991,
  author = {Affleck, Ian and Ludwig, Andreas W. W.},
  title = {Universal Noninteger ``Ground-State Degeneracy'' in Critical Quantum Systems},
  journal = {Phys. Rev. Lett.},
  volume = {67},
  pages = {161--164},
  year = {1991},
  doi = {10.1103/PhysRevLett.67.161},
  url = {https://doi.org/10.1103/PhysRevLett.67.161}
}

@article{AffleckLudwig1993,
  author = {Affleck, Ian and Ludwig, Andreas W. W.},
  title = {Exact Conformal-Field-Theory Results on the Multichannel Kondo Effect: Single-Fermion Green's Function, Self-Energy, and Resistivity},
  journal = {Phys. Rev. B},
  volume = {48},
  pages = {7297--7321},
  year = {1993},
  doi = {10.1103/PhysRevB.48.7297},
  url = {https://doi.org/10.1103/PhysRevB.48.7297}
}

@article{Affleck1995,
  author = {Affleck, Ian},
  title = {Conformal Field Theory Approach to the Kondo Effect},
  journal = {Acta Phys. Pol. B},
  volume = {26},
  pages = {1869--1932},
  year = {1995},
  eprint = {cond-mat/9512099},
  archivePrefix = {arXiv},
  url = {https://www.actaphys.uj.edu.pl/R/26/12/1869}
}

@incollection{Cardy2004,
  author = {Cardy, John L.},
  title = {Boundary Conformal Field Theory},
  booktitle = {Encyclopedia of Mathematical Physics},
  pages = {333--340},
  publisher = {Elsevier},
  year = {2006},
  doi = {10.1016/B0-12-512666-2/00398-9},
  eprint = {hep-th/0411189},
  archivePrefix = {arXiv},
  url = {https://doi.org/10.1016/B0-12-512666-2/00398-9}
}

@article{NozieresBlandin1980,
  author = {Nozi{\`e}res, Philippe and Blandin, Andr{\'e}},
  title = {Kondo Effect in Real Metals},
  journal = {J. Phys. France},
  volume = {41},
  pages = {193--211},
  year = {1980},
  doi = {10.1051/jphys:01980004103019300},
  url = {https://doi.org/10.1051/jphys:01980004103019300}
}

@article{CoxZawadowski1998,
  author = {Cox, D. L. and Zawadowski, A.},
  title = {Exotic Kondo Effects in Metals: Magnetic Ions in a Crystalline Electric Field and Tunnelling Centres},
  journal = {Adv. Phys.},
  volume = {47},
  pages = {599--942},
  year = {1998},
  doi = {10.1080/000187398243500},
  url = {https://doi.org/10.1080/000187398243500}
}

@article{KaneFisher1992,
  author = {Kane, C. L. and Fisher, Matthew P. A.},
  title = {Transport in a One-Channel Luttinger Liquid},
  journal = {Phys. Rev. Lett.},
  volume = {68},
  pages = {1220--1223},
  year = {1992},
  doi = {10.1103/PhysRevLett.68.1220},
  url = {https://doi.org/10.1103/PhysRevLett.68.1220}
}

@article{MandelstamTamm1945,
  author = {Mandelstam, L. and Tamm, I.},
  title = {The Uncertainty Relation between Energy and Time in Non-Relativistic Quantum Mechanics},
  journal = {J. Phys. (USSR)},
  volume = {9},
  pages = {249--254},
  year = {1945},
  url = {https://cir.nii.ac.jp/crid/1370285709897301766}
}

@article{Hasegawa2023,
  author = {Hasegawa, Yoshihiko},
  title = {Unifying Speed Limit, Thermodynamic Uncertainty Relation and Heisenberg Principle via Bulk-Boundary Correspondence},
  journal = {Nat. Commun.},
  volume = {14},
  pages = {2828},
  year = {2023},
  doi = {10.1038/s41467-023-38074-8},
  url = {https://doi.org/10.1038/s41467-023-38074-8}
}

@article{Funo2025,
  author = {Funo, Ken and Tajima, Hiroyasu},
  title = {Symmetry Induced Enhancement in Finite-Time Thermodynamic Trade-Off Relations},
  journal = {Phys. Rev. Lett.},
  volume = {134},
  pages = {080401},
  year = {2025},
  doi = {10.1103/PhysRevLett.134.080401},
  url = {https://doi.org/10.1103/PhysRevLett.134.080401}
}

@article{Yamauchi2025,
  author = {Yamauchi, Aoi and Nagase, Rihito and Li, Kaixin and Sagawa, Takahiro and Funo, Ken},
  title = {Thermodynamic Speed Limit for Non-Adiabatic Work and Its Classical--Quantum Decomposition},
  journal = {J. Phys. A: Math. Theor.},
  volume = {58},
  pages = {205001},
  year = {2025},
  doi = {10.1088/1751-8121/add4d5},
  eprint = {2502.19661},
  archivePrefix = {arXiv},
  primaryClass = {quant-ph},
  url = {https://doi.org/10.1088/1751-8121/add4d5}
}

@misc{SM,
  note = {See Supplemental Material at [URL] for detailed evaluation.}
}

@article{Naze2023,
  title = {Optimal work fluctuations for finite-time and weak processes},
  author = {Naz\'e, Pierre},
  journal = {Phys. Rev. E},
  volume = {108},
  issue = {5},
  pages = {054118},
  numpages = {4},
  year = {2023},
  month = {Nov},
  publisher = {American Physical Society},
  doi = {10.1103/PhysRevE.108.054118},
  url = {https://link.aps.org/doi/10.1103/PhysRevE.108.054118}
}

@book{FractionalCalculus,
  author    = {Samko, S. G. and Kilbas, A. A. and Marichev, O. I.},
  title     = {Fractional Integrals and Derivatives: Theory and Applications},
  publisher = {Gordon and Breach Science Publishers},
  address   = {Yverdon},
  year      = {1993},
  isbn      = {978-2-88124-864-1},
  url = {https://ci.nii.ac.jp/ncid/BA21828325}
}

@book{Gradshteyn2014,
  author    = {Gradshteyn, I. S. and Ryzhik, I. M.},
  editor    = {Zwillinger, Daniel and Moll, Victor H.},
  title     = {Table of Integrals, Series, and Products},
  edition   = {8},
  publisher = {Academic Press},
  address   = {Amsterdam},
  year      = {2014},
  isbn      = {978-0-12-384933-5},
  url       = {https://www.sciencedirect.com/book/9780123849335/table-of-integrals-series-and-products}
}

@misc{DLMF,
  key          = {{\relax DLMF}},
  title        = {{\it NIST Digital Library of Mathematical Functions}},
  howpublished = {\url{https://dlmf.nist.gov/}, Release 1.2.6 of 2026-03-15},
  url          = {https://dlmf.nist.gov/},
  note         = {F.~W.~J. Olver, A.~B. {Olde Daalhuis}, D.~W. Lozier, B.~I. Schneider, R.~F. Boisvert, C.~W. Clark, B.~R. Miller, B.~V. Saunders, H.~S. Cohl, and M.~A. McClain, eds.}
}

@book{Rainville1960,
  author    = {Rainville, Earl D.},
  title     = {Special Functions},
  publisher = {The Macmillan Company},
  address   = {New York},
  year      = {1960},
  url = {https://ci.nii.ac.jp/ncid/BA10939277}
}

@misc{Forrester2026,
  author        = {Forrester, Peter J.},
  title         = {Gegenbauer polynomials and fluctuation properties of the one-dimensional {Riesz} gas},
  year          = {2026},
  month         = {April},
  eprint        = {2604.25078},
  archivePrefix = {arXiv},
  primaryClass  = {math-ph},
  doi           = {10.48550/arXiv.2604.25078},
  url           = {https://arxiv.org/abs/2604.25078},
  note          = {arXiv:2604.25078}
}

@article{Beato2025,
  title = {Toward a Theory of Phase Transitions in Quantum Control Landscapes},
  author = {Beato, Nicol\`o and Patil, Pranay and Bukov, Marin},
  journal = {Phys. Rev. X},
  volume = {15},
  issue = {4},
  pages = {041014},
  numpages = {42},
  year = {2025},
  month = {Oct},
  publisher = {American Physical Society},
  doi = {10.1103/27nn-cb3t},
  url = {https://link.aps.org/doi/10.1103/27nn-cb3t}
}

@mastersthesis{Karush1939,
  author = {Karush, William},
  title  = {Minima of Functions of Several Variables with Inequalities as Side Constraints},
  school = {Department of Mathematics, University of Chicago},
  year   = {1939},
  url = {https://books.google.com/books/about/Minima_of_Functions_of_Several_Variables.html?id=o5NmHQAACAAJ}
}

@inproceedings{KuhnTucker1951,
  author    = {Kuhn, Harold W. and Tucker, Albert W.},
  title     = {Nonlinear Programming},
  booktitle = {Proceedings of the Second Berkeley Symposium on Mathematical Statistics and Probability},
  pages     = {481--492},
  publisher = {University of California Press},
  address   = {Berkeley},
  year      = {1951},
  doi = {10.1525/9780520411586-036},
  url = {https://doi.org/10.1525/9780520411586-036}
}

@book{BoydVandenberghe2004,
  author    = {Boyd, Stephen and Vandenberghe, Lieven},
  title     = {Convex Optimization},
  publisher = {Cambridge University Press},
  address   = {Cambridge},
  year      = {2004},
  doi = {10.1017/CBO9780511804441},
  url = {https://doi.org/10.1017/CBO9780511804441}
}

@article{Iftikhar2015,
  author  = {Iftikhar, Z. and Jezouin, S. and Anthore, A. and Gennser, U. and Parmentier, F. D. and Cavanna, A. and Pierre, F.},
  title   = {Two-channel {Kondo} effect and renormalization flow with macroscopic quantum charge states},
  journal = {Nature},
  volume  = {526},
  number  = {7572},
  pages   = {233--236},
  year    = {2015},
  doi     = {10.1038/nature15384},
  url = {https://doi.org/10.1038/nature15384}
}

@article{Iftikhar2018,
  author  = {Iftikhar, Z. and Anthore, A. and Mitchell, A. K. and Parmentier, F. D. and Gennser, U. and Ouerghi, A. and Cavanna, A. and Mora, C. and Simon, P. and Pierre, F.},
  title   = {Tunable quantum criticality and super-ballistic transport in a ``charge'' {Kondo} circuit},
  journal = {Science},
  volume  = {360},
  number  = {6395},
  pages   = {1315--1320},
  year    = {2018},
  doi     = {10.1126/science.aan5592},
  url = {https://doi.org/10.1126/science.aan5592}
}

\section*{End Matter}

\emsection{Derivation of the Fredholm equation}Let $v(t)=\dot{g}(t)$ and impose the endpoint constraint $\int_0^\tau v(t) \mathrm{d} t=1$, minimizing Eq. (\ref{eq:Var}) with a Lagrange multiplier $\chi$ gives
\begin{equation}
\delta\left[\alpha^2\langle v, \hat{K} v\rangle-2 \chi\left(\int_0^\tau v(t) \mathrm{d} t-1\right)\right]=0,
\end{equation}
where $(\hat{K} v)(t)=\int_0^\tau \mathcal{C}\left(t-t^{\prime}\right) v\left(t^{\prime}\right) \mathrm{d} t^{\prime}$. Hence
\begin{equation}
\hat{K} v^*=\frac{\chi}{\alpha^2} 1 . 
\end{equation}
Writing $\zeta=\hat{K}^{-1} 1$ gives the optimal driven speed $v^*=\zeta / Z$, with $Z=\langle 1, \zeta\rangle$. The minimum is
\begin{equation}
\sigma^2_{\scriptscriptstyle W\mathrm{op}}=\alpha^2\left\langle v^*, \hat{K} v^*\right\rangle=\frac{\alpha^2}{Z}.
\end{equation}

\emsection{Finite-temperature conformal map}Starting from the fluctuation Fredholm equation, $\int_0^\tau \mathcal{C}(t-t',\beta)\zeta(t')\dd t'=1$, introduce
\begin{equation}
\mu=\tanh\!\left[\pi\left(t-\frac{\tau}{2}\right)/\beta\right],
\quad
\nu=\tanh\!\left[\pi\left(t'-\frac{\tau}{2}\right)/\beta\right].
\end{equation}
Both variables lie in $[-a,a]$, where $a=\tanh(\pi \tau/2\beta)$. The change of variables gives $\dd t'=\dd\nu/[\pi (1-\nu^2)/\beta]$. Using $\tanh A-\tanh B=\sinh(A-B)/(\cosh A\cosh B)$, one obtains
\begin{equation}
|\mu-\nu|=|\sinh[\pi (t-t')/\beta]|
\sqrt{(1-\mu^2)(1-\nu^2)} .
\end{equation}
Therefore
\begin{equation}
\frac{1}{|\sinh[\pi (t-t')/\beta]|^{2\Delta}}
=\frac{(1-\mu^2)^\Delta(1-\nu^2)^\Delta}
{|\mu-\nu|^{2\Delta}} .
\end{equation}
Substituting this relation into the Fredholm equation and rescale $\nu=ax$ and $\mu=ay$ yields
\begin{equation}
\begin{aligned}
1={}&\beta^2C_\Delta(\frac{\pi }{\beta})^{2\Delta-1}
a^{1-2\Delta}(1-a^2y^2)^\Delta \\
&\times\int_{-1}^{1}\dd x\,
\frac{(1-a^2x^2)^{\Delta-1}\zeta[t(x)]}
{|y-x|^{2\Delta}} .
\end{aligned}
\label{eq:em-scaled-equation}
\end{equation}
Defining
\begin{equation}
u(x)=\left(\frac{\pi }{a\beta}\right)^{2\Delta-1}
(1-a^2x^2)^{\Delta-1}\zeta[t(x)]
\end{equation}
turns Eq.~\eqref{eq:em-scaled-equation} into Eq.~\eqref{eq:carleman-main}. The inverse relation is exactly Eq.~\eqref{eq:inverse-map-main}. This derivation shows explicitly why temperature appears only through $a$ after the conformal transformation.

\emsection{Asymptotic solutions of speed limit}The minimum fluctuation at finite temperature is \cite{SM}
\begin{equation}
\sigma^2_{\scriptscriptstyle W \mathrm{op}}(\tau,\beta)=
\frac{\pi\hbar\alpha^{2}(\pi /\beta)^{2\Delta}}
{\Lambda^{2\Delta}B\left(\frac12,\Delta+\frac12\right)
a^{2\Delta}{}_{2}F_{1}(1,\Delta;1+\Delta;a^{2})}.
\label{eq:SuppFiniteTVar}
\end{equation}
For the marginal case $\Delta=1/2$, since $a=\tanh(\pi \tau/2\beta)$, we obtain $B\left(\frac12,1\right)a\,{}_{2}F_{1}\left(1,\frac12;\frac32;a^2\right)=\pi\tau/\beta$. Substituting into Eq.~\eqref{eq:SuppFiniteTVar} yields
\begin{equation}
  \sigma^2_{\scriptscriptstyle W \mathrm{op}}(\tau,\beta)^{\Delta=1/2}
    =
    \frac{\pi\hbar\alpha^{2}}{\Lambda\tau}.
\end{equation}
Therefore, imposing $\sigma^2_{\scriptscriptstyle W \mathrm{op}}\leq\sigma^2_{\scriptscriptstyle W}$ gives the speed limit for $\Delta=1/2$ shown in Eq. (\ref{eq:C2CK}).

For the limit $\tau \gg\beta$ corresponds to  $a\to 1 $. In this regime, the zero-balanced hypergeometric function has the asymptotic form
\begin{align*}
{}_{2}F_{1}(1,\Delta;1+\Delta;a^{2})&=
\Delta\left[-\log(1-a^{2})-\gamma-\psi(\Delta)\right]\\
&+O\!\left((1-a^{2})\log(1-a^{2})\right).    
\end{align*}
where \(\gamma\) is the Euler constant and \(\psi(z)\) is the digamma function. Since $1-a^{2}=\operatorname{sech}^{2}(\pi \tau/2\beta)$, we have $-\log(1-a^{2})=\pi \tau/\beta-\log 4+O(e^{-\pi \tau/\beta})$. Therefore, to leading order, ${}_{2}F_{1}(1,\Delta;1+\Delta;a^{2})\approx\Delta\,\pi \tau/\beta$  with $a^{2\Delta}\to 1 $. Substituting this into Eq.~\eqref{eq:SuppFiniteTVar} gives
\begin{equation}
\sigma^2_{\scriptscriptstyle W \mathrm{op}}(\tau,\beta)=\frac{\hbar\alpha^{2}\pi^{2\Delta}}{\Lambda^{2\Delta}\Delta B\left(\frac12,\Delta+\frac12\right)}\frac{\beta^{1-2\Delta}}{\tau}.   
\end{equation}

Inverting this expression at fixed tolerance gives the speed limit for adiabatic limit:
\begin{equation}
\tau^{\AL}_{\mathrm{SL}}=\frac{\pi\hbar\alpha^{2}(\pi T)^{2\Delta-1}}
{\Lambda^{2\Delta}\Delta
B\left(\frac12,\Delta+\frac12\right)\sigma^2_{\scriptscriptstyle W}}.
\label{eq:SLAL}
\end{equation}

For the limit \(\tau\ll\beta\), namely \(a\to0\), the approximation \(a=\tanh(\pi \tau/2\beta)\approx \pi \tau/2\beta\) gives
\[
a^{2\Delta}{}_{2}F_{1}(1,\Delta;1+\Delta;a^{2})=\left({\pi \tau}/{2\beta}\right)^{2\Delta}+O(\left({\pi \tau}/{2\beta}\right)^{2\Delta+2}).
\]

Substituting this into Eq.~\eqref{eq:SuppFiniteTVar} and using $2^{2\Delta}/B\left(\frac12,\Delta+\frac12\right)=1/
B\left(\Delta+\frac12,\Delta+\frac12\right)$, we have
\begin{equation}
  \sigma^2_{\scriptscriptstyle W \mathrm{op}}(\tau,\beta)^{\mathrm{KZ}}
=
\frac{\pi\hbar\alpha^{2}}
{(\Lambda\tau)^{2\Delta}
B\left(\Delta+\frac12,\Delta+\frac12\right)}.
\end{equation}

Inverting this expression gives the Kibble-Zurek speed limit shown in Eq.~\eqref{eq:SLKZ}.

\emsection{Asymptotic solutions of the transformed Fredholm equation for optimal protocol}The two limits used in the main text are obtained by expanding the source term in Eq.~\eqref{eq:carleman-main} before acting with the inverse Riesz operator $\hat{\mathcal K}_\Delta^{-1}$, where $(\hat{\mathcal K}_\Delta u)(y)=\int_{-1}^1u(x)|y-x|^{-2\Delta}\dd x$.

For $\tau \gg\beta$, write $\epsilon=1-a^2=\sech^2(\pi \tau/2\beta)$. We can expand  $(1-a^2y^2)^{-\Delta}$  in series, thus
\[
\begin{aligned}
u(x)={}&\lambda_\Delta\hat{\mathcal K}_\Delta^{-1}
\!\left[(1-y^2)^{-\Delta}\right](x) \\
&-\Delta\epsilon\lambda_\Delta\hat{\mathcal K}_\Delta^{-1}
\!\left[\frac{y^2}{(1-y^2)^{\Delta+1}}\right](x)
+O(\epsilon^2) .
\end{aligned}
\]
The leading inverse is \cite{FractionalCalculus}
\begin{equation}
\hat{\mathcal K}_\Delta^{-1}\!\left[(1-y^2)^{-\Delta}\right](x)
=\frac{(1-x^2)^{\Delta-1}}
{2^{2\Delta}B(1-2\Delta,\Delta)} .
\end{equation}
Inserting this result into Eq.~\eqref{eq:inverse-map-main} and using the identity $2^{2\Delta}B(1-2\Delta,\Delta)=\pi/(\Delta \cos(\pi\Delta)B(\frac12,\Delta+\frac12))$, we obtain the leading inverse-kernel profile
\begin{equation}
\zeta_{\AL}(t)=
\frac{\Lambda^{2\Delta}\Delta B(\frac12,\Delta+\frac12)}
{\pi\hbar^{2\Delta}\beta^{3-2\Delta}} .
\end{equation}
Since this expression is independent of $t$, the linear protocol is the optimal protocol in the adiabatic limit. The same expression also gives Eq.~\eqref{eq:SLAL}.

For $\tau \ll\beta$, the source is instead analytic in $a^2$, consequently,
\[
u(x)=\lambda_\Delta\hat{\mathcal K}_\Delta^{-1}[1](x)
+\Delta a^2\lambda_\Delta\hat{\mathcal K}_\Delta^{-1}[y^2](x)
+O(a^4) .
\]
The leading inverse is most transparently obtained from the Gegenbauer eigenfunction relation for the Riesz kernel on $[-1,1]$,
\begin{equation}
\hat{\mathcal K}_\Delta^{-1}[1](x)
=\frac{\cos(\pi\Delta)}{\pi}(1-x^2)^{\Delta-1/2} .
\end{equation}
Therefore the leading mapped solution is
\begin{equation}
u_{\KZ}(x)=
\frac{\Lambda^{2\Delta}}{\pi\beta^2\hbar}
(1-x^2)^{\Delta-1/2} .
\end{equation}
Using $a\simeq\pi \tau/2\beta$ and $x\simeq2t/\tau-1$ in Eq.~\eqref{eq:inverse-map-main} gives
\begin{equation}
\zeta_{\KZ}(t)=
\frac{\Lambda^{2\Delta}}{\pi\beta^2\hbar}
[t(\tau-t)]^{\Delta-1/2}.
\end{equation}
Together with $\int_0^\tau[t(\tau-t)]^{\Delta-1/2}\dd t=\tau^{2\Delta}B(\Delta+1/2,\Delta+1/2)$, this yields Eqs.~\eqref{eq:kz-var-protocol} and \eqref{eq:SLKZ}.

\end{document}


\title{Supplementary material: Work fluctuation speed limit in boundary conformal field theories}
\author{Shihao Xia, Ahsan Nazir, Harry J. D. Miller}
\affiliation{Department of Physics and Astronomy, The University of Manchester,
Manchester M13 9PL, United Kingdom}
\date{\today}

\maketitle
\tableofcontents
\renewcommand{\theequation}{S\arabic{equation}}
\renewcommand{\thetable}{S\arabic{table}}
\renewcommand{\thefigure}{S\arabic{figure}}
\renewcommand{\thesection}{S\arabic{section}}
\bibliographystyle{apsrev4-1}

\section{Kernel function in time space}

\subsection{Derivation of $O_{k}(t)$}

In frequency space, the kernel function of the cumulants takes the
form:\cite{Guarnieri2024}:

\begin{equation}
\beta^{k}\kappa_{w}^{k}=\int_{\mathbb{R}}\frac{d\omega}{2\pi}\tilde{\Psi}_{0}(\omega)\gamma^{k}(\omega)\left|\int_{0}^{\tau}dt\dot{\lambda}_{t}e^{i\omega t}\right|^{2}\geq0,\label{eq:cw}
\end{equation}

where 

\begin{equation}
\gamma^{k}(\omega):=\begin{cases}
\frac{1}{2}(\beta\hbar\omega)^{k-1}\textrm{Coth}(\beta\hbar\omega/2) & \text{ if }k\text{ even }\\
\frac{1}{2}(\beta\hbar\omega)^{k-1} & \text{ if }k\text{ odd. }
\end{cases}\label{eq:gamma}
\end{equation}
Eq. (\ref{eq:cw}) can be further calculated as follows:

\begin{equation}
\begin{aligned}\beta^{k}\kappa_{W}^{k} & =\int_{0}^{\tau}dt\int_{0}^{\tau}dt\dot{\lambda}_{t}\dot{\lambda}_{t^{\prime}}\left[\mathcal{F}^{-1}\gamma^{k}(\omega)*\tilde{\Psi}_{0}(\omega)\right]\left(t-t^{\prime}\right)\\
 & =\int_{0}^{\tau}dt\int_{0}^{\tau}dt\dot{\lambda}_{t}\dot{\lambda}_{t^{\prime}}O_{k}\left(t-t^{\prime}\right),
\end{aligned}
\end{equation}

where the kernel function $O_{k}\left(t-t^{\prime}\right)$ is defined
as

\begin{equation}
O_{k}\left(t-t^{\prime}\right)=\int_{-\infty}^{+\infty}dv\mathcal{F}^{-1}[\gamma^{k}(\omega)](v)\Psi_{0}\left(t-t^{\prime}-v\right).\label{eq:Kernel}
\end{equation}

To obtain the formula of $O_{k}\left(t-t^{\prime}\right)$, I first
calculated the inverse Fourier transform of $\gamma^{k}(\omega)$
, for odd $k$ $(k=2m-1,m=1,2,3...)$, I have

\begin{equation}
\begin{aligned}\mathcal{F}^{-1}[r^{k}(\omega)](t) & =\int_{\mathbb{R}}\frac{d\omega}{2\pi}e^{-i\omega t}\frac{1}{2}(\beta\hbar\omega)^{k-1}\\
 & =\frac{(\beta\hbar)^{2m-2}}{2}\int_{\mathbb{R}}\frac{d\omega}{2\pi}(-i\omega)^{2m-2}e^{-i\omega t}(-1)^{m-1}\\
 & =\frac{(i\beta\hbar)^{2m-2}}{2}\frac{d^{2m-2}}{dt^{2m-2}}\delta(t)
\end{aligned}
\label{eq:Fodd}
\end{equation}

Substitute Eq. (\ref{eq:Fodd}) into Eq. (\ref{eq:Kernel}), we obtain:

\begin{equation}
\begin{aligned}O_{k\textrm{odd}}\left(t-t^{\prime}\right) & =\int_{\mathbb{R}}dv\frac{(i\beta\hbar)^{2m-2}}{2}\frac{d^{2m-2}}{dv^{2m-2}}\delta(v)\Psi_{0}\left(t-t^{\prime}-v\right)\\
 & =\frac{(i\beta\hbar)^{2m-2}}{2}\Psi_{0}^{2m-2}\left(t-t^{\prime}\right),
\end{aligned}
\label{eq:Kodd}
\end{equation}

where the convolution property of the delta function, $\int_{R}\delta^{(n)}(v)f(t-v)dv=f^{(n)}(t)$,
was used.

Next, I calculate the inverse Fourier transform of $\gamma^{k}(\omega)$
for $k$ even $(k=2m,m=1,2,3...)$, yielding:

\begin{equation}
\begin{aligned}\mathcal{F}^{-1}[r^{k}(\omega)](t) & =\int_{\mathbb{R}}\frac{d\omega}{2\pi}e^{-i\omega t}\frac{(\beta\hbar\omega)^{k-1}}{2}\textrm{Coth}\left(\frac{\beta\hbar}{2}\omega\right)\\
 & =\frac{(i\beta\hbar)^{2m-1}}{2}\frac{d^{2m-1}}{dv^{2m-1}}\int_{\mathbb{R}}\frac{d\omega}{2\pi}e^{-i\omega t}\textrm{Coth}\left(\frac{\beta\hbar}{2}\omega\right)\\
 & =\frac{(i\beta\hbar)^{2m-2}}{2}\frac{d^{2m-1}}{dv^{2m-1}}\textrm{Coth}\left(\frac{\pi t}{\beta\hbar}\right).
\end{aligned}
\label{eq:Feven}
\end{equation}

Substituting Eq. (\ref{eq:Feven}) into Eq. (\ref{eq:Kernel}), we
have:

\begin{equation}
\begin{aligned}
O_{k\textrm{even}}\left(t-t^{\prime}\right) & =\frac{(i\beta\hbar)^{2m-2}}{2}\int_{\mathbb{R}}dv\mathrm{\frac{d^{2m-1}}{dv^{2m-1}}Coth}\left(\frac{\pi v}{\beta\hbar}\right)\Psi_{0}\left(t-t^{\prime}-v\right)\\
& =\frac{(i\beta\hbar)^{2m-2}}{2}\int_{\mathbb{R}}dv\mathrm{Coth}^{(2m-1)}\left(\frac{\pi v}{\beta\hbar}\right)\Psi_{0}\left(t-t^{\prime}-v\right)\\
 & =\frac{(i\beta\hbar)^{2m-2}}{2}\int_{\mathbb{R}}dv\mathrm{Coth}\left(\frac{\pi v}{\beta\hbar}\right)\Psi_{0}^{(2m-1)}\left(t-t^{\prime}-v\right).
\end{aligned}
\label{eq:Keven}
\end{equation}

The last line we used integral by part and the property $\Psi_{0}\left(\tau\rightarrow\infty\right)=0$.
Finally, we arrive at:

\begin{equation}
\begin{aligned}O_{k}(\tau) & =\begin{cases}
\frac{(i\beta\hbar)^{2m-2}}{2}\Psi_{0}^{2m-2}(\tau) & k=2m-1\\
\frac{(i\beta\hbar)^{2m-2}}{2}\int_{\mathbb{R}}dv\textrm{Coth}^{(2m-1)}\left(\frac{\pi v}{\beta\hbar}\right)\Psi_{0}\left(\tau-v\right) & k=2m
\end{cases}\end{aligned}
.\label{eq:KernelFin}
\end{equation}

where $m=1,2,3...$ Next we test the parity of $O_{k}(\tau)$. Since
the parity of $\Psi_{0}\left(\tau\right)$ is even, $O_{k}$ is also
even for odd $k$. For even $k$, we have

\begin{equation}
\begin{aligned}O_{k}(-\tau) & =\frac{(i\beta\hbar)^{2m-2}}{2}\int_{\mathbb{-\infty}}^{+\infty}dv\textrm{Coth}^{(2m-1)}\left(\frac{\pi v}{\beta\hbar}\right)\Psi_{0}\left(-\tau-v\right)\\
 & =\frac{(i\beta\hbar)^{2m-2}}{2}\int_{\mathbb{-\infty}}^{+\infty}du\textrm{Coth}^{(2m-1)}\left(\frac{\pi u}{\beta\hbar}\right)\Psi_{0}\left(\tau-u\right)\\
 & =O_{k}(\tau).
\end{aligned}
\label{eq:Kevenparity}
\end{equation}

where the parities of $\Psi_{0}\left(\tau\right)$ (even) and $\textrm{Coth}\left(\tau\right)$
(odd) are have been used. This proves the parity of the function $O_{k}(\tau)$
for any $k$. 

Then we try to reduce to the classical situation to check our result.
First, we consider the high-temperature limit, $\beta\hbar\rightarrow0$.
Using the approximation $\textrm{Coth}(\frac{\pi v}{\beta\hbar})\approx\textrm{sgn}(\frac{\pi v}{\beta\hbar})$ and
noting that $\frac{d}{dx}\textrm{sgn}(x)=2\delta(x)$, we obtain:

\begin{equation}
O_{2}(\tau)\approx\frac{1}{2}\int_{\mathbb{R}}dv\dot{\textrm{sgn}}\left(\frac{\pi v}{\beta\hbar}\right)\Psi_{0}\left(\tau-v\right)=\int_{\mathbb{R}}dv\delta\left(\frac{\pi v}{\beta\hbar}\right)\Psi_{0}\left(\tau-v\right)=\Psi_{0}\left(\tau\right)=2O_{1}(\tau).
\end{equation}

Furthermore, considering the fact that $[H_{0},V]=0$, which implies
$\Psi_{0}(\tau)=\Psi_{0}(0)$, we have:

\begin{equation}
O_{2}(\tau)=\frac{1}{2}\Psi_{0}(\tau)\int_{\mathbb{R}}dv\dot{\textrm{Coth}}\left(\frac{\pi v}{\beta\hbar}\right)=\Psi_{0}\left(0\right)\left.\textrm{Coth}\left(\frac{\pi v}{\beta\hbar}\right)\right|_{-\infty}^{+\infty}=2\Psi_{0}\left(0\right)=2O_{1}(\tau).
\end{equation}

These relations show that

\begin{equation}
\beta\left\langle W_{\textrm{diss}}\right\rangle =\frac{1}{2}\beta^{2}\textrm{Var}(W_{\textrm{diss}}).
\end{equation}

\subsection{Time reversal symmetry for optimal protocol}

The Euler-Lagrange equation is used to optimize the $k-th$ order
of the CGF\cite{Naze2023}, we can calculate the kernel function of the
cumulants with respect to time $t$ as:

\begin{equation}
\beta^{k}\kappa_{w}^{k}=\alpha^{2}\int_{0}^{\tau}dt\int_{0}^{\tau}dt^{\prime}\dot{g}(t)\dot{g}\left(t^{\prime}\right)O_{k}\left(t-t^{\prime}\right),\label{eq:ct}
\end{equation}

where we represent $\lambda_{t}=\alpha g(t)$ and set the boundary
conditions as $g(0)=0$ and $g(\tau)=1$. Optimize Eq. (\ref{eq:ct})
is the main thing in this paper. Integrating by parts yields:

\begin{equation}
\begin{aligned}\beta^{k}k_{w}^{k} & =\alpha^{2}\left(O_{k}(0)-\int_{0}^{\tau}dO_{k}(t-\tau)g(t)-\int_{0}^{\tau}d^{\prime}O_{k}\left(\tau-t^{\prime}\right)g(t')-\int_{0}^{\tau}\int_{0}^{\tau}dtdt^{\prime}\ddot{O}_{k}\left(t-t^{\prime}\right)g(t)g(t')\right)\\
 & =\alpha^{2}\left(O_{k}(0)+2\int_{0}^{\tau}dt\dot{O}_{k}(\tau-t)g(t)-\int_{0}^{\tau}\int_{0}^{\tau}dtdt^{\prime}\ddot{O}_{k}\left(t-t^{\prime}\right)g(t)g(t')\right),
\end{aligned}
\label{eq:cumbypart}
\end{equation}

where the parity of $O_{k}$ was used in the second line. Next, we
calculate the variations of Eq.(\ref{eq:cumbypart}). Setting $\beta^{k}k_{w}^{k}=I_{0}+I_{1}+I_{2}$,
we have $\delta I_{0}=0$ ,

\begin{equation}
\begin{aligned}\delta I_{1} & =2\alpha^{2}\int_{0}^{\tau}dt\dot{O}_{k}(\tau-t)\delta g(t)\end{aligned}
,\label{eq:Var1st}
\end{equation}

and

\begin{equation}
\begin{aligned}\delta I_{2}= & -2\alpha^{2}\int_{0}^{\tau}\int_{0}^{\tau}dtdt^{\prime}\ddot{O}_{k}\left(t-t^{\prime}\right)g\left(t^{\prime}\right)\delta g(t),\end{aligned}
\label{eq:Var2nd}
\end{equation}

where the parity of $O_{k}$ is used here. Combining Eq.(\ref{eq:Var1st})
and Eq.(\ref{eq:Var2nd}), and applying Euler-Lagrange equation, the
optimal control protocol $g_{k}^{*}\left(t^{\prime}\right)$ for any
$k-th$ order of CGF is found to be:

\begin{equation}
\dot{O}_{k}(\tau-t)=\int_{0}^{\tau}dt^{\prime}\ddot{O}_{k}\left(t-t^{\prime}\right)g_{k}^{*}\left(t^{\prime}\right).\label{eq:Optimal}
\end{equation}

The relation between reverse and forward processes is $\lambda_{R}(t)=\lambda_{F}(\tau-t)=\lambda_{0}+g_{F}(\tau-t)\delta\lambda.$
Additionally, any reverse process that links the same equilibrium
states as $\lambda_{F}(t)$ can be written as

\begin{equation}
\begin{aligned}\lambda_{R}(t) & =\lambda_{\tau}-g_{R}(t)\delta\lambda\\
 & =\lambda_{0}+g_{F}(\tau)\delta\lambda-g_{R}(t)\delta\lambda\\
 & =\lambda_{0}+\left(g_{F}(\tau)-g_{R}(t)\right)\delta\lambda\\
 & =\lambda_{0}+\left(1-g_{R}(t)\right)\delta\lambda
\end{aligned}
\end{equation}

Thus, we have $g_{F}(\tau-t)=1-g_{R}(t),$or

\begin{equation}
g_{F}(t)=1-g_{R}(\tau-t).\label{eq:TRS}
\end{equation}

Substituting Eq.(\ref{eq:TRS}) into Eq.(\ref{eq:Optimal}), we have

\begin{equation}
\begin{aligned} & \int_{0}^{\tau}dt^{\prime}\ddot{O}_{k}\left(t-t^{\prime}\right)\left(1-g^{*}(\tau-t^{\prime})\right)dt^{\prime}\\
 & =\int_{0}^{\tau}dt^{\prime}\ddot{O}_{k}(t-t^{\prime})-\int_{0}^{\tau}dt^{\prime}\ddot{O}_{k}(t-t^{\prime})g^{*}(\tau-t^{\prime})\\
 & =\dot{O}_{k}(t)-\dot{O}_{k}(t-\tau)-\int_{0}^{\tau}dt^{\prime\prime}\ddot{O}_{k}(t-\tau+t^{\prime\prime})g^{*}(t^{\prime\prime})\\
 & =\dot{O}_{k}(\tau-t)
\end{aligned}
\end{equation}

By changing the variable $t\rightarrow\tau-t$ and using the parity
of $O_{k},$we get

\begin{equation}
\int_{0}^{\tau}dt^{\prime\prime}\ddot{O}_{k}\left(\tau-t^{\prime\prime}\right)g^{*}\left(t^{\prime\prime}\right)=\dot{O}_{k}\left(\tau-t\right).
\end{equation}

This result indicates that both $g_{k}^{*}(t)$ and its inverse $1-g_{k}^{*}(\tau-t)$
are optimal protocols. This relation holds for any order $k$ and
their linear combinations. For example, we can built the kernel function
as:

\begin{equation}
I=a\beta\left\langle W_{diss}\right\rangle +(1-a)\beta^{2}\textrm{Var}\left(W_{diss}\right),\label{eq:Ia}
\end{equation}

its optimal protocol obeys to this symmetry for any $a$.

\section{Physical interpretation and derivation for Finite Temperature Results}

\subsection{Conformal structure of $  \mathcal{C}(t,\beta)$}

To evaluate $  \mathcal{C}(t,\beta)$, which equivalent to $O_{2}(t,\beta)$,  and reveal its underlying physical structure,
it is highly effective to map it to the frequency domain using the
expansion. We introduce the spectral function shown in Eq. (A27) in
\cite{Guarnieri2024}
\begin{equation}
\tilde{\Psi}_{0}(\omega)=2\pi\frac{\beta(1-e^{-\beta\hbar\omega})}{\hbar\omega}\sum_{n,m}\delta\left(\omega-\beta^{-1}\log(p_{n}/p_{m})\right)\delta V_{nm}p_{m}\delta V_{mn}.
\end{equation}

By taking the Fourier transform of $  \mathcal{C}(t,\beta)$ and combining it with
this spectral representation, the operator can be expressed as 
\begin{equation}
\widetilde{\mathcal{C}}(\omega,\beta)=2\pi\frac{1}{2}(\beta\hbar\omega)\coth\left(\frac{\beta\hbar\omega}{2}\right)\Psi_{0}(\omega).
\end{equation}

Expanding the hyperbolic cotangent, this simplifies to 
\begin{equation}
\begin{aligned}\widetilde{\mathcal{C}}(\omega,\beta) & =2\pi\cdot\frac{1}{2}(\beta\hbar\omega)\textrm{coth}\left(\frac{\beta\hbar\omega}{2}\right)\Psi_{0}(\omega)\\
 & =\pi\beta^{2}\left(1+e^{-\beta\hbar\omega}\right)\sum_{n,m}\delta\left(\omega-\beta^{-1}\log\left(p_{n}/p_{m}\right)\right)\delta V_{nm}p_{m}\delta V_{mn}.
\end{aligned}
\end{equation}

This specific frequency-domain structure naturally connects to the
two-point correlation function of the system, defined as 
\begin{equation}
S(t)=\langle\delta V(t)\delta V(0)\rangle=\mathrm{Tr}\left[\pi_{\beta}\delta V(t)\delta V(0)\right]=\sum_{n,m}p_{n}\delta V_{nm}e^{i(E_{n}-E_{m})t/\hbar}\delta V_{mn}.
\end{equation}

Its Fourier transform is given by 
\begin{equation}
\tilde{S}(\omega)=\int_{\mathbb{R}}dte^{i\omega t}S(t)=2\pi\sum_{n,m}p_{n}\delta V_{nm}\delta V_{mn}\delta\left(\omega-\log(p_{n}/p_{m})/\beta\hbar\right).
\end{equation}

By exchanging the summation indices and exploiting the thermal weights,
one straightforwardly recovers the Kubo-Martin-Schwinger (KMS) relation
in the frequency domain, 
\begin{equation}
\tilde{S}(-\omega)=e^{-\beta\hbar\omega}\tilde{S}(\omega)
\end{equation}

Concurrently, in the time domain, $S(-\omega)=\int_{\mathbb{R}}d(-t)e^{i\omega(-t)}S(-t),$
and the corresponding KMS relation ensures that $S(-t)=\langle\delta V(-t)\delta V(0)\rangle=\langle\delta V(0)\delta V(t)\rangle$.
Substituting these symmetric properties back into our expression for
$\widetilde{\mathcal{C}}(\omega,\beta)$, we obtain 
\begin{equation}
\widetilde{\mathcal{C}}(\omega,\beta)=\frac{1}{2}\beta^{2}\left(\widetilde{S}(\omega)+\widetilde{S}(-\omega)\right).
\end{equation}

Transforming this result back to the time domain reveals a crucial
physical identification: $  \mathcal{C}(t,\beta)$ is exactly proportional to the
anti-commutator of the perturbation operators, giving 
\begin{equation}
  \mathcal{C}(t,\beta)=\frac{1}{2}\beta^{2}\bigl(S(t)+S(-t)\bigr)=\frac{1}{2}\beta^{2}\langle\{\delta V(t),\delta V(0)\}\rangle_0.
\end{equation}

where $\{a,b\}$ means anti-commutator. This quantity is fixed entirely
by conformal symmetry. From Refs.\cite{Ma2025,AffleckLudwig1993}, we know
that $\langle\delta V(t),\delta V(0)\rangle_0\propto|t|^{-2\Delta}$.
Applying the cylinder mapping and restoring the scale-dependent coefficient
$C_{\Delta}$, the corresponding finite-temperature kernel can be
written in the analytic form 
\begin{equation}
\mathcal{C}(t,\beta)=\beta^2C_{\Delta}\left[\frac{\pi/\beta}{|\sinh (\pi (t-i0^+)/\beta)|}\right]^{2\Delta}.\label{eq:ConCov}
\end{equation}

\subsection{Evaluation of $  \mathcal{C}(t,\beta)$}

In this part we derive a closed-form expression for $  \mathcal{C}(t,\beta)$. We
begin with the result in \cite{Ma2025}

\[
\Psi_{0}(t,\beta)=A_{\Delta}\int_{0}^{\textrm{csch}^{2}(\pi t/\beta)}\frac{v^{\Delta-1}}{\sqrt{1+v}}dv,
\]

where $A_{\Delta}=\frac{\pi^{2\Delta-1}{\beta}^{2-2\Delta}}{\Lambda^{2\Delta}}\sin(\pi\Delta)$. For convenience, here we set $\hbar=1$.
After integrating the original representation Eq.(\ref{eq:KernelFin})
by parts, one may write

\begin{equation}
  \mathcal{C}(t,\beta)=\frac{1}{2}\int_{\mathbb{R}}dv\coth(\pi v/\beta)\partial_{t}\Psi_{0}(t-v,\beta).\label{eq:O2conv}
\end{equation}

The boundary term vanishes because ${\Psi_{0}(t-v)\to0}$
as ${v\to\pm\infty}$. To evaluate the time derivative,
it is convenient to denote $y(t)=\textrm{csch}^{2}(\pi t/\beta)$, for
which $y'(t)=-\frac{2\pi }{\beta}\coth(\pi t/\beta)\textrm{csch}^{2}(\pi t/\beta)$. Using
the fundamental theorem of calculus, together with $1+\textrm{csch}^{2}x=\coth^{2}x$
and hence $\sqrt{1+\textrm{csch}^{2}x}=|\coth x|$, we obtain

\begin{equation}
    \partial_{t}\Psi_{0}(t)=-\frac{2\pi }{\beta} A_{\Delta}\mathrm{sgn}(t)\bigl[\textrm{csch}^{2}(\pi t/\beta)\bigr]^{\Delta}.\label{dPsi}
\end{equation}

Substituting this result into Eq. (\ref{eq:O2conv}) and changing
variables from $v$ to $u=t-v$, we find

\[
  \mathcal{C}(t,\beta)=-A_{\Delta}\frac{\pi }{\beta}\int_{\mathbb{R}}du\coth\bigl(\pi (t-u)/\beta\bigr)\mathrm{sgn}(u)\bigl[\textrm{csch}(\pi |u|/\beta)\bigr]^{2\Delta}.
\]

Separating the regions $u>0$ and $u<0$ gives

\begin{equation}
  \mathcal{C}(t,\beta)=-A_{\Delta}\frac{\pi }{\beta}\int_{0}^{\infty}du\bigl[\textrm{csch}(\pi u/\beta)\bigr]^{2\Delta}\left[\coth\bigl(\pi (t-u)/\beta\bigr)-\coth\bigl(\pi (t+u)/\beta\bigr)\right].\label{eq:O2split}
\end{equation}

The bracket can be simplified by the standard identity ${\coth(a-b)-\coth(a+b)=2\cosh b\sinh b/(\sinh^{2}a-\sinh^{2}b)}.$
With $a=\pi t/\beta$ and $b=\pi u/\beta$, Eq. (\ref{eq:O2split}) becomes

\[
  \mathcal{C}(t,\beta)=-A_{\Delta}\frac{\pi }{\beta}\int_{0}^{\infty}du\frac{2\cosh(\pi u/\beta)\sinh(\pi u/\beta)^{1-2\Delta}}{\sinh^{2}(\pi t/\beta)-\sinh^{2}(\pi u/\beta)}.
\]

Next, introducing the dimensionless variable $s=\sinh^{2}(\pi u/\beta)/\sinh^{2}(\pi t/\beta)$,
so that ${ds=2\pi T\frac{\sinh(\pi u/\beta)\cosh(\pi u/\beta)}{\sinh^{2}(\pi t/\beta)}du},$
the integral reduces to

\[
  \mathcal{C}(t,\beta)=-A_{\Delta}\frac{1}{\sinh^{2\Delta}(\pi t/\beta)}\mathrm{P.v.}\int_{0}^{\infty}ds\frac{s^{-\Delta}}{1-s}.
\]

Here the principal value is required because the integrand has a simple
pole at $s=1$. For $0<\Delta<1/2$, the standard formula $\mathrm{P.v.}\int_{0}^{\infty}dss^{-\Delta}/(1-s)=-\pi\cot(\pi\Delta)$, see 3.241.3 in \cite{Gradshteyn2014}. Adding the $\hbar$ back and substituting the coefficient $A_{\Delta}$
yields

\[
  \mathcal{C}(t,\beta)=\frac{\beta^{2} \hbar \cos(\pi\Delta)}{\Lambda^{2\Delta}}\left(\frac{\pi /\beta}{\sinh(\pi t/\beta)}\right)^{2\Delta},
\]

which is Eq.(8) we shown in main article.

\subsection{The marginal case $\Delta=1/2$}

The endpoint $\Delta=1/2$ has to be treated separately.  In this
case the regular principal-value part of the time-domain kernel
vanishes, while a contact term remains.  We derive it most directly
in frequency space.  For $\Delta=1/2$,
\[
A_{1/2}=\frac{\beta}{\Lambda},
\]
and thus the Eq. (\ref{dPsi}) is
\[
\partial_{t}\Psi_{0}(t)
=
-\frac{2\pi}{\Lambda}
\operatorname{sgn}(t)
\operatorname{csch}(\pi  |t|/\beta).
\]
Then we can calculate
\begin{equation}
\begin{aligned}
\widetilde{\Psi}_0^{\prime}(\omega)&=-\frac{2 \pi}{\Lambda } \int_{-\infty}^{\infty} d t e^{i \omega t} \operatorname{sgn}(t) \operatorname{csch}(\pi |t|/\beta) \\
&=-\frac{4 \pi i}{\Lambda} \int_0^{\infty} d t \frac{\sin (\omega t)}{\sinh (\pi t/\beta)}.
\end{aligned}
\end{equation}
Using the result 3.981.1 in \cite{Gradshteyn2014}. 
\[
\int_0^\infty dt\,
\frac{\sin(\omega t)}{\sinh(a t)}
=
\frac{\pi}{2a}
\tanh\frac{\pi\omega}{2a},
\]
with $a=\pi /\beta$. Notice that $\widetilde\Psi_0(\omega)=-\widetilde{\Psi}_0^{\prime}(\omega)/i\omega$ and combine with $\hbar$ give
\[
\widetilde\Psi_0(\omega)
=
\frac{2\pi\beta}{\Lambda \hbar\omega}
\tanh\frac{\hbar\omega\beta}{2}.
\]
For the second cumulant,
\[
\gamma_2(\omega)
=
\frac{\beta\hbar\omega}{2}
\coth\frac{\beta\hbar\omega}{2}.
\]
Therefore
\[
\widetilde{\mathcal{C}}(\omega,\beta)
=
\widetilde\Psi_0(\omega)\gamma_2(\omega)
=
\frac{\pi\beta^2\hbar}{\Lambda}.
\]
Thus the marginal kernel is local in time,
\[
  \mathcal{C}(t,\beta)=\frac{\pi\hbar\beta^2}{\Lambda}\delta(t).
\]
The Fredholm equation therefore reduces to
\[
\frac{\pi\beta^2\hbar}{\Lambda}\zeta(t)=1,
\]
so that
\[
\zeta(t)=\frac{\Lambda}{\pi\beta^2\hbar},
\qquad
g_2^*(t)=\frac{t}{\tau}.
\]
The optimized fluctuation is
\[\mathrm{Var}(W)_{\mathrm{op}}=\frac{\pi\hbar\alpha^2}{\Lambda\tau}.
\]

\subsection{Finite-temperature optimal value}

Here we derive the finite-temperature optimal value of the work variance from the Fredholm problem. The calculation is included mainly to justify the compact expression used in the main text. The optimal protocol itself requires the pointwise inverse kernel and is not needed for the scalar minimum.

In the main article, we introduced the conformal mapped Fredholm equation which reduces to the finite-interval Riesz equation
\begin{equation}
(\hat{\mathcal K}_{\Delta}u)(x)
=\frac{1}{\beta^{2}C_{\Delta}}f_{a}(x),
\label{eq:SuppRieszEq}
\end{equation}
where $(\hat{\mathcal K}_{\Delta}u)(x)=
\int_{-1}^{1}\frac{u(y)}{|x-y|^{2\Delta}}dy,$ and $f_{a}(x)=(1-a^{2}x^{2})^{-\Delta}.$ If we set $q_{a}=\hat{\mathcal K}_{\Delta}^{-1}f_{a}$ , then $u=(\beta^{2}C_{\Delta})^{-1}q_{a}$ and
\begin{equation}
Z_{\tau}=\left(\frac{a \beta}{\pi}\right)^{2\Delta}
\frac{1}{\beta^{2}C_{\Delta}}
\mathcal J_{\Delta}(a),
\qquad
\mathcal J_{\Delta}(a)
=\left\langle f_{a},\hat{\mathcal K}_{\Delta}^{-1}f_{a}\right\rangle .
\label{eq:SuppJdef}
\end{equation}
Therefore
\begin{equation}
\sigma^2_{\scriptscriptstyle W \mathrm{op}}(\tau,\beta)
=\frac{\alpha^{2}C_{\Delta}\pi^{2\Delta}}
{a^{2\Delta}\mathcal J_{\Delta}(a)\beta^{2\Delta}}.
\label{eq:SuppVarBeforeJ}
\end{equation}
It remains to evaluate the scalar quadratic form $\mathcal J_{\Delta}$.

We use the standard Gegenbauer diagonalization of the finite-interval
Riesz kernel \cite{Forrester2026,Rainville1960,DLMF,FractionalCalculus}. With the weighting function
\begin{equation}
w_{\Delta}(x)=(1-x^{2})^{\Delta-1/2},
\end{equation}
one has
\begin{equation}
\hat{\mathcal K}_{\Delta}
\left[w_{\Delta}(x)C_{n}^{\Delta}(x)\right](y)
=\lambda_{n}C_{n}^{\Delta}(y),
\qquad
\lambda_{n}=\frac{\pi}{\cos(\pi\Delta)}\frac{(2\Delta)_{n}}{n!}.
\label{eq:SuppGegenEigen}
\end{equation}
Together with the orthogonality relation
\begin{equation}
\int_{-1}^{1}w_{\Delta}(x)C_{n}^{\Delta}(x)C_{m}^{\Delta}(x)dx
=h_{n}^{\Delta}\delta_{nm},
\qquad
h_{n}^{\Delta}=\frac{\pi2^{1-2\Delta}\Gamma(n+2\Delta)}
{n!(n+\Delta)\Gamma^{2}(\Delta)},
\end{equation}
this gives the inverse-kernel representation
\begin{equation}
\hat{\mathcal K}_{\Delta}^{-1}f(x)
=w_{\Delta}(x)\sum_{n=0}^{\infty}\frac{f_{n}}{\lambda_{n}}C_{n}^{\Delta}(x),
\end{equation}
Simultaneously, we can also expand $f(x)$  by Gegenbauer polynomials as  $f_a(x)=\sum_{n=0}^{\infty}f_{n}(a)C_{n}^{\Delta}(x).$ As a result,
\begin{equation}
\mathcal J_{\Delta}(a)=
\sum_{n=0}^{\infty}\frac{h_{n}^{\Delta}}{\lambda_{n}}f_{n}^{2}(a).
\label{eq:SuppJspectral}
\end{equation}
Since $f_{a}(x)$ is even, only even Gegenbauer modes contribute, which means, $f_{2m+1}(a)=0$. Expanding
$(1-a^{2}x^{2})^{-\Delta}$ using the binomial expansion $\sum_{r=0}^{\infty} \frac{(\Delta)_r}{r!} a^{2r} x^{2 r}$ and monomial-to-Gegenbauer formula
\cite{Rainville1960,Forrester2026}:

\begin{equation}
x^p=\frac{p!}{2^p} \sum_{k=0}^{[p / 2]} \frac{\nu+p-2 k}{k!(\nu)_{p+1-k}} C_{p-2 k}^{(\nu)}(x),
\end{equation}

and also combining the properties of rising factorials yields
\begin{equation}
f_{2m}(a)=
a^{2m}\frac{(\Delta)_{m}(1/2)_{m}}{(\Delta)_{2m}}
\,{}_{2}F_{1}\left(m+\Delta,m+\frac12;2m+\Delta+1;a^{2}\right).
\label{eq:Suppf2m}
\end{equation}
Substitution into Eq. (\ref{eq:SuppJspectral}) gives
\begin{equation}
\mathcal J_{\Delta}(a)=G_{\Delta}
\sum_{m=0}^{\infty}\frac{a^{4m}}{2m+\Delta}
\left[\frac{(\Delta)_{m}(1/2)_{m}}{(\Delta)_{2m}}\right]^{2}
\left[{}_{2}F_{1}\left(m+\Delta,m+\frac12;2m+\Delta+1;a^{2}\right)\right]^{2},
\label{eq:SuppJseries}
\end{equation}
where $G_{\Delta}=\cos(\pi\Delta)2^{1-2\Delta}\Gamma(2\Delta)/\Gamma^{2}(\Delta).$ The remaining summation can be carried out by standard hypergeometric identities. First, the square of the Gauss function is converted by Clausen's identity \cite{Vidunas_2011},
\begin{equation}
\left[{}_{2}F_{1}(a,b;a+b+1/2;z)\right]^{2}
={}_{3}F_{2}(2a,2b,a+b;2a+2b,a+b+1/2;z).
\end{equation}
Writing $z=a^{2}$ and extracting the coefficient of $z^{N}$, one obtains
\begin{equation}
\mathcal J_{\Delta}(a)=G_{\Delta}\sum_{N=0}^{\infty}S_{N}z^{N},
\qquad
S_{N}=\sum_{m=0}^{\lfloor N/2\rfloor}T_{m,N}.
\end{equation}
Combining with the properties of rising factorials, we found that the ratio of the general term to the $m=0$ term is
\begin{equation}
\frac{T_{m,N}}{T_{0,N}}=
\frac{(\Delta)_{m}(1+\Delta/2)_{m}(1/2)_{m}(-N/2)_{m}((1-N)/2)_{m}}
{(\Delta/2)_{m}(\Delta+1/2)_{m}((N+2\Delta+1)/2)_{m}((N+2\Delta+2)/2)_{m}m!},
\label{eq:SuppTermRatio}
\end{equation}
with
\begin{equation}
T_{0,N}=\frac{1}{\Delta}
\frac{(2\Delta)_{N}(\Delta+1/2)_{N}}
{(2\Delta+1)_{N}(\Delta+1)_{N}}.
\end{equation}
The parameters $-N/2$ and $(1-N)/2$ ensure that the corresponding
${}_{5}F_{4}(1)$ series terminates at $m=\lfloor N/2\rfloor$. Applying
the terminating Rogers--Dougall summation formula \cite{DLMF} 

\begin{equation}
\begin{aligned}
{ }_5 F_4\left(
a, \frac{1}{2} a+1, b, c, d ;
\frac{1}{2} a, a-b+1, a-c+1, a-d+1
; 1\right)\\=\frac{\Gamma(a-b+1) \Gamma(a-c+1) \Gamma(a-d+1) \Gamma(a-b-c-d+1)}{\Gamma(a+1) \Gamma(a-b-c+1) \Gamma(a-b-d+1) \Gamma(a-c-d+1)},    
\end{aligned}
\end{equation}

gives
\begin{equation}
S_{N}=\frac{1}{N+\Delta}.
\label{eq:SuppSN}
\end{equation}
Hence
\begin{equation}
\mathcal J_{\Delta}(a)=G_{\Delta}
\sum_{N=0}^{\infty}\frac{(a^{2})^{N}}{N+\Delta}
=\frac{G_{\Delta}}{\Delta}
{}_{2}F_{1}(1,\Delta;1+\Delta;a^{2}).
\end{equation}
Using the duplication formula for the Gamma function, the prefactor can
be written as
\begin{equation}
\frac{G_{\Delta}}{\Delta}
=\frac{\cos(\pi\Delta)}{\pi}B\left(\frac12,\Delta+\frac12\right),
\end{equation}
and therefore
\begin{equation}
\mathcal J_{\Delta}(a)
=\frac{\cos(\pi\Delta)}{\pi}B\left(\frac12,\Delta+\frac12\right)
{}_{2}F_{1}(1,\Delta;1+\Delta;a^{2}) .
\label{eq:SuppJclosed}
\end{equation}
Combining Eqs. (\ref{eq:SuppVarBeforeJ}) and (\ref{eq:SuppJclosed})
finally gives the finite-temperature optimal variance
\begin{equation}
\sigma^2_{\scriptscriptstyle W \mathrm{op}}(\tau,\beta)=
\frac{\hbar\pi\alpha^{2}(\pi T)^{2\Delta}}
{\Lambda^{2\Delta}B\left(\frac12,\Delta+\frac12\right)
a^{2\Delta}{}_{2}F_{1}(1,\Delta;1+\Delta;a^{2})}.
\label{eq:SuppFiniteTVar}
\end{equation}
Equivalently,
\begin{equation}
a^{2\Delta}{}_{2}F_{1}(1,\Delta;1+\Delta;a^{2})=\Delta B_{a^2}(\Delta,0)
=\Delta\int_{0}^{a^{2}}\frac{s^{\Delta-1}}{1-s}ds,
\end{equation}
which makes the monotonicity in $a$. Therefore for a fluctuation $\sigma^2_{\scriptscriptstyle W}$, Eq. (\ref{eq:SuppFiniteTVar}) gives the finite-temperature speed limit
\begin{equation}
\tau\geq\tau_{SL}=\frac{2}{\pi T}\operatorname{arctanh}
\left[\mathcal F^{-1}\left(
\frac{\hbar\pi\alpha^{2}(\pi T)^{2\Delta}}
{\Lambda^{2\Delta}B\left(\frac12,\Delta+\frac12\right)\sigma^2_{\scriptscriptstyle W}}
\right)\right],
\label{eq:SuppFiniteTSL}
\end{equation}
where $\mathcal F_\Delta(a)=a^{2\Delta}{}_{2}F_{1}(1,\Delta;1+\Delta;a^{2})=\Delta B_{a^2}(\Delta,0)$, equivalently an incomplete-beta function, and $\mathcal F^{-1}$ is its inverse on $0<a^2<1$. 

\section{Boundary-CFT Scaling of Protocol Timescales}

\subsection{Finite Temperature scaling effect}

The logarithmic linearity follows from the scaling structure of the two timescales.  For the linear protocol, the change of variables \(u=\pi T s\) gives 
\begin{equation}
\sigma^2_{\scriptscriptstyle W}=K_\Delta( T/\Lambda)^{2\Delta}H_\Delta(\pi T\tau_{\rm lin}),     
\end{equation}
where  \(K_\Delta=2\hbar\alpha^2\cos(\pi\Delta)\pi^{2\Delta}\) and \(H_\Delta(x)=x^{-2}\int_0^x{\rm d}u\,(x-u)\sinh^{-2\Delta}u\). Thus \(\Lambda\tau_{\rm lin}=(\pi  T/\Lambda)^{-1}
H_\Delta^{-1}[(K_\Delta q)^{-1}]\), with \(q=(T/\Lambda)^{2\Delta}/\sigma^2_{\scriptscriptstyle W}\).  The speed-limit time has the same scaling form, \(\Lambda\tau_{\rm SL}=(\pi  T/\Lambda)^{-1}S_\Delta(q)\), where $S_\Delta(q)$ is defined by Eq. (\ref{eq:SuppFiniteTSL}), because its argument also depends on \(T/\Lambda\) and \(\sigma^2_{\scriptscriptstyle W}\) only through \((T/\Lambda)^{2\Delta}/\sigma^2_{\scriptscriptstyle W}\). Consequently, the ratio \(\tau_{\rm SL}/\tau_{\rm lin}\) is a function only of \(q\).  Fixing \(\tau_{\rm SL}/\tau_{\rm lin}=R\) therefore fixes \(q=q_R\) on a given branch, and hence

\begin{equation}
\log_{10}(T/\Lambda)
=
\frac{1}{2\Delta}\log_{10}\sigma^2_{\scriptscriptstyle W}
+
\frac{1}{2\Delta}\log_{10}q_R .
\end{equation}

Therefore, at fixed \(\tau_{\rm SL}/\tau_{\rm lin}\), the curve in the \((\log_{10}\sigma^2_{\scriptscriptstyle W},\log_{10}(T/\Lambda)\) plane is a straight line with slope \(1/(2\Delta)\).  

\subsection{Velocity-constrained speed limit for Kibble-Zurek regime}

To treat the velocity-constrained problem, we first remove the explicit time scale from the optimization problem.  In the KZ regime, the variance functional has the homogeneous form
\begin{equation}
\sigma^2_{\scriptscriptstyle W}[g]
=
C_\Delta
\frac{\hbar\alpha^2}{\Lambda^{2\Delta}}
\int_0^\tau dt
\int_0^\tau dt'\,
\frac{\dot g(t)\dot g(t')}{|t-t'|^{2\Delta}} .
\end{equation}
Introducing the rescaled time \(s=t/\tau\) and the dimensionless velocity \(w(s)=\tau \dot g(t)\), the boundary condition becomes \(\int_0^1 w(s)\,ds=1\), while the velocity constraint \(0\leq \dot g(t)\leq v_{\max}\) becomes \(0\leq w(s)\leq M\), with
\begin{equation}
M=\tau v_{\max}.    
\end{equation}
Therefore the constrained optimization problem is naturally written as the dimensionless problem
\begin{equation}
    J(M)
=
\min_{w}
\int_0^1 ds
\int_0^1 ds'\,
\frac{w(s)w(s')}{|s-s'|^{2\Delta}},
\end{equation}
subject to $\int_0^1 w(s)ds=1$ and $0\leq w(s)\leq M$. The important point is that \(J(M)\) depends only on the scaled bound \(M\) and on \(\Delta\), but not on the cutoff \(\Lambda\).  The unconstrained speed-limit result corresponds to \(M\to\infty\), and we denote \(J_0=\lim_{M\to\infty}J(M)\).

After the rescaling, the constrained variance takes the form
\begin{equation}
\sigma^2_{\scriptscriptstyle W}(\tau,M)
=
A\tau^{-2\Delta}\frac{J(M)}{J_0},
\qquad
A=
\frac{\pi\hbar\alpha^2}
{\Lambda^{2\Delta}
B\left(\Delta+\frac12,\Delta+\frac12\right)} .
\end{equation}
For the same target variance \(\sigma^2_{\scriptscriptstyle W}\), the unconstrained KZ speed-limit time satisfies
\begin{equation}
\sigma^2_{\scriptscriptstyle W}=A\left(\tau_{u}\right)^{-2\Delta}.    
\end{equation}
Comparing the constrained and unconstrained expressions gives
\begin{equation}
\frac{\tau_{u}}{\tau_c(M)}
=
\left(\frac{J_0}{J(M)}\right)^{1/(2\Delta)}
\equiv R(M).    
\end{equation}
Thus the ratio \(\tau_{u}/\tau_c\) is a function only of the dimensionless bound \(M\).  In particular, it has no explicit dependence on \(\Lambda\).

The physical velocity is recovered from \(M=\tau v_{\max}\).  Since \(\tau_c(M)=\tau_{u}/R(M)\), we obtain
\begin{equation}
v_{\max}(M)
=
\frac{M}{\tau_c(M)}
=
\frac{M R(M)}{\tau_{u}} .    
\end{equation}
Using
\begin{equation}
\tau_{u}
=
\frac{1}{\Lambda}
\left[
\frac{\pi\hbar\alpha^2}
{B\left(\Delta+\frac12,\Delta+\frac12\right)\sigma^2_{\scriptscriptstyle W}}
\right]^{1/(2\Delta)},    
\end{equation}
one finds
\begin{equation}
\frac{v_{\max}(M)}{\Lambda}
=
M R(M)
\left[
\frac{
B\left(\Delta+\frac12,\Delta+\frac12\right)\sigma^2_{\scriptscriptstyle W}
}
{\pi\hbar\alpha^2}
\right]^{1/(2\Delta)} .    
\end{equation}
Equivalently,  for fixed $M$
\begin{equation}
\frac{\tau_{u}}{\tau_c(M)}
=\frac{v_{\max}(M)}{\Lambda M}
\left[
\frac{\pi\hbar\alpha^2}
{B\left(\Delta+\frac12,\Delta+\frac12\right)\sigma^2_{\scriptscriptstyle W}}
\right]^{1/(2\Delta)}\propto\frac{v_{\max}(M)}{\Lambda}.    
\end{equation}

The right-hand side is independent of \(\Lambda\).  Hence, after solving the scaled optimization problem, the natural physical horizontal variable is  \(v_{\max}/\Lambda\).  

\bibliographystyle{apsrev4-2}
\bibliography{refs}